\documentclass[12pt,french,psamsfonts]{amsart}

\usepackage[french]{babel}
\usepackage{amsmath}
\usepackage{amsthm}
\usepackage{amssymb}
\usepackage{amscd}
\usepackage{amsfonts}
\usepackage{amsbsy}
\usepackage{epsfig,afterpage}
\usepackage{psfrag}

\input cyracc.def
\font\tencyr=wncyr10
\def\cyr{\tencyr\cyracc}

\begin{document}

\title[Blondel et les oscillations auto-entretenues]{Blondel et les oscillations auto-entretenues}

\author[J.M. Ginoux $^1$ {\&} R. Lozi $^2$]
{Jean-Marc Ginoux$^1$ {\&} Ren\'{e} Lozi$^2$}

\address{$^1$ Laboratoire {\sc Protee}, I.U.T. de Toulon,
Universit\'{e} du Sud, BP 20132, F-83957 La Garde Cedex, France}
\email{ginoux@univ-tln.fr}

\address{$^2$ Laboratoire {\sc Laboratoire J.A. Dieudonn\'{e}},
Universit\'{e} de Nice Sophia-Antipolis, Parc Valrose, 06108 NICE Cedex 02, France}
\email{lozi@unice.fr}

\maketitle



\begin{abstract}

En 1893, \og physicien-ing\'{e}nieur \fg{} Andr\'{e} Blondel invente l'oscillographe bifilaire permettant de visualiser les tensions et courants variables. \`{A} l'aide de ce puissant moyen d'investigation, il entreprend tout d'abord l'\'{e}tude des ph\'{e}nom\`{e}nes de l'arc \'{e}lectrique alors utilis\'{e} pour l'\'{e}clairage c\^{o}tier et urbain puis de l'arc chantant employ\'{e} comme \'{e}metteur d'ondes radio\'{e}lectriques en T.S.F. En 1905, il met en \'{e}vidence un nouveau type d'oscillations non-sinusoïdales au sein de l'arc chantant. Vingt ans plus tard, Balthazar Van der Pol reconnaitra qu'il s'agissait en r\'{e}alit\'{e} d'oscillations de relaxation. Pour expliquer ce ph\'{e}nom\`{e}ne, il fait appel \`{a} une representation dans le plan de phase et montre que son \'{e}volution prend la forme de petits cycles. Pendant le premier conflit mondial la triode remplace peu \`{a} peu l'arc chantant dans les syst\`{e}mes de transmission. Au sortir de la guerre, Blondel transpose \`{a} la triode la plupart de ses r\'{e}sultats par analogie avec l'arc chantant. En avril 1919, il publie un long m\'{e}moire dans lequel il introduit la terminologie \og oscillations auto-entretenues \fg{} et propose d'illustrer ce concept \`{a} partir de l'exemple du vase de Tantale qui sera ensuite repris par Van der Pol et par Philippe Le Corbeiller. Il fournit alors une d\'{e}finition d'un syst\`{e}me auto-entretenu assez proche de celles qui seront donn\'{e}es par la suite par Aleksandr' Andronov et Van der Pol. Pour \'{e}tudier la stabilit\'{e} des oscillations entretenues par la triode et par l'arc chantant il utilise cette fois une repr\'{e}sentation dans le plan complexe et explicite l'amplitude en coordonn\'{e}es polaires. Il justifie alors l'entretien des oscillations par l'existence de cycles qui pr\'{e}sentent presque toutes les caract\'{e}ristiques des cycles limites de Poincar\'{e}. Enfin, en novembre 1919, Blondel r\'{e}alise, un an avant Van der Pol, la mise en \'{e}quation des oscillations d'une triode. En mars 1926, Blondel \'{e}tablit l'\'{e}quation diff\'{e}rentielle caract\'{e}risant les oscillations de l'arc chantant, en tous points similaire \`{a} celle qu'obtient concomitamment Van der Pol pour la triode. Ainsi, tout au long de sa carri\`{e}re, Blondel, a apport\'{e} une contribution fondamentale et relativement m\'{e}connue \`{a} l'\'{e}laboration de la th\'{e}orie des oscillations non lin\'{e}aires. L'objet de cet article est donc d'analyser ses principaux travaux dans ce domaine et de mesurer leur importance, voire leur influence en les replaçant dans la perspective du d\'{e}veloppement de cette th\'{e}orie.

\end{abstract}

\newpage

\section{Introduction}

\`{A} la fin du XIX$^{e}$ si\`{e}cle physiciens et ing\'{e}nieurs observent un ph\'{e}nom\`{e}ne oscillatoire d'un genre nouveau dans diff\'{e}rents dispositifs comme les machines s\'{e}rie-dynamo \cite{Lescuyer}, les machines hydrauliques munies d'un r\'{e}gulateur \`{a} action indirecte \cite{Léauté} ou les machines comportant des arcs \'{e}lectriques\footnote{\`{A} cette \'{e}poque c'\'{e}tait le cas de la plupart des syst\`{e}mes d'\'{e}clairage.}. Ils cherchent tout d'abord \`{a} \'{e}liminer ces oscillations qu'ils qualifient alors d'entretenues\footnote{Cette terminologie englobait \`{a} la fois les oscillations forc\'{e}es et les oscillations auto-entretenues.} et consid\`{e}rent comme d\'{e}l\'{e}t\`{e}res avant de r\'{e}aliser ensuite toute leur importance.\\

En France, la v\'{e}tust\'{e} des installations \'{e}lectriques des syst\`{e}mes de signalisation maritime incite le jeune ing\'{e}nieur Andr\'{e} Blondel (1863-1938), affect\'{e} en 1889 au Service central des Phares et Balises, \`{a} effectuer des recherches sur l'\textit{arc \'{e}lectrique} \`{a} courants alternatifs dans le but d'am\'{e}liorer ce genre de dispositif.

\begin{quote}
\og En 1889, les connaissances sur les propri\'{e}t\'{e}s des courants alternatifs \'{e}taient tr\`{e}s limit\'{e}es et leur \'{e}tude \'{e}tait arr\^{e}t\'{e}e par l'absence de moyens suffisamment pratiques d'analyse exp\'{e}rimentale. Charg\'{e} d'\'{e}tudier la production et l'emploi des arc \'{e}lectriques \`{a} courants alternatifs pour les phares, j'ai donc d\^{u}  cr\'{e}er les instruments de travail n\'{e}cessaire. \fg{} \cite[p. IX]{Blondel1911}
\end{quote}

C'est ainsi que Blondel \cite{Blondel1893a} invente en 1893 l'oscillographe bifilaire permettant l'inscription directe  et instantan\'{e}e des tensions ou des courants variables qui sera ensuite am\'{e}lior\'{e} par Duddell \cite{Duddell1897}, puis peu \`{a} peu remplac\'{e} par l'oscilloscope cathodique de Braun \cite{Braun}. Gr\^{a}ce \`{a} ce puissant moyen d'investigation, Blondel put d\'{e}terminer toutes les propri\'{e}t\'{e}s de l'\textit{arc \'{e}lectrique} \`{a} courants alternatifs et permit, d'apr\`{e}s Joseph Bethenod \cite[p. 751]{Bethenod1938} qui fut son assistant de 1904 \`{a} 1907, de \og  faire accomplir un pas d\'{e}cisif \`{a} la th\'{e}orie de l'arc \'{e}lectrique \fg{}.\\

\`{A} Londres, l'ing\'{e}nieur britannique William Du Bois Duddell (1872-1917) fut mandat\'{e}, en 1899, par les autorit\'{e}s anglaises pour trouver une solution au probl\`{e}me du bruit engendr\'{e} par le syst\`{e}me d'\'{e}clairage des rues utilisant des lampes \`{a} arc \'{e}lectrique. En plaçant en d\'{e}rivation un circuit oscillant comportant une bobine et un condensateur aux bornes de l'arc \'{e}lectrique il parvint \`{a} faire cesser le bruissement.

\newpage

Apr\`{e}s avoir ensuite constat\'{e} que pour certaines valeurs de la capacit\'{e} et de l'inductance, ce dispositif \'{e}mettait des sons audibles pour l'homme, Duddell \cite{Duddell1900a,Duddell1900b,Duddell1901} lui donna le nom de \textit{musical arc}\footnote{En France, il semble que ce soit l'ing\'{e}nieur Paul Janet (1863-1937) qui ait traduit cette expression par \textit{arc chantant} \cite{Janet1902}.} et r\'{e}alisa tout l'int\'{e}r\^{e}t que ce dispositif pr\'{e}sentait pour la T\'{e}l\'{e}graphie Sans Fil\footnote{Il semble que d\`{e}s 1898 Blondel ait envisag\'{e} la possibilit\'{e} d'une telle application. Voir pli cachet\'{e} n$^\circ$ 6041, Blondel \cite{Blondel1913}.}. En effet, on utilisait jusqu'alors dans le domaine de la T\'{e}l\'{e}graphie Sans Fil (T.S.F.) l'\textit{oscillateur de Hertz}. L'\'{e}tincelle \'{e}lectrique qui \og \'{e}clatait \fg{} \\entre deux sph\`{e}res de cuivres engendrait une onde \'{e}lectromagn\'{e}tique susceptible d'\^{e}tre interpr\'{e}t\'{e}e par un r\'{e}cepteur comme un signal sous forme d'impulsions. L'un des inconv\'{e}nients majeurs de ce dispositif \'{e}tait que l'oscillation \'{e}lectrique produite, c'est-\`{a}-dire, le signal \'{e}tait tr\`{e}s rapidement amorti. L'\textit{arc chantant} allait en revanche permettre d'entretenir ces oscillations donnant ainsi un essor consid\'{e}rable \`{a} la T.S.F.\\

Entre 1897 et 1905, Andr\'{e} Blondel va plublier plusieurs articles afin \og d'\'{e}lucider compl\`{e}tement les
ph\'{e}nom\`{e}nes de l'\textit{arc chantant}. \fg{} \cite[p. 752]{Bethenod1938}. Tout d'abord, Blondel \cite[p. 519]{Blondel1897} d\'{e}montre, contrairement \`{a} une hypoth\`{e}se alors tr\`{e}s r\'{e}pandue, l'absence de \textit{force contre-\'{e}lectromotrice} au sein de l'arc. En juillet 1905, Blondel \cite{Blondel1905c} pr\'{e}sente devant les membres de la Soci\'{e}t\'{e} française de Physique un imposant m\'{e}moire \og Sur les ph\'{e}nom\`{e}nes de l'arc chantant \fg{} dans lequel il met en \'{e}vidence un nouveau type \textit{discontinu} d'oscillations pour ce dispositif qu'il qualifie d'\textit{arc sifflant}. Afin d'analyser la stabilit\'{e} de ce r\'{e}gime oscillatoire il introduit la notion de \textit{caract\'{e}ristique d'oscillation} qui consiste \`{a} repr\'{e}senter son \'{e}volution dans le plan de phase et montre qu'elle prend la forme de petits cycles ce qui le conduit \`{a} fournir une premi\`{e}re d\'{e}finition du concept d'\textit{auto-entretien}.\\

Trois ans plus tard, lors d'une s\'{e}rie de conf\'{e}rences (r\'{e}cemment re\-d\'{e}couvertes\footnote{Voir \`{a} ce sujet la th\`{e}se de Ginoux \cite{Gith}.}) effectu\'{e}es en mai-juin 1908 \`{a} l'\'{E}cole Sup\'{e}rieure Professionnelle des Postes et T\'{e}l\'{e}graphes o\`{u} il enseigne depuis 1902, Henri Poincar\'{e} \cite{P6} est l'un des premiers en France\footnote{L'\'{E}cole Allemande repr\'{e}sent\'{e}e par Theodor Hermann Simon \cite{Simon1906} et Heinrich Barkha\"{u}sen \cite{Barkhausen} semble y \^{e}tre parvenue concomitamment.} \`{a} \'{e}tablir l'\'{e}quation incompl\`{e}te\footnote{Dans la mesure o\`{u} il ne fournit pas l'expression de la fonction non lin\'{e}aire repr\'{e}sentant la \textit{f.\'{e}.m.} de l'arc.} des oscillations entretenues par un \textit{arc chantant}.

\newpage

Il parvient n\'{e}anmoins \`{a} d\'{e}montrer (vingt ans avant Andronov \cite{Andronov1928,Andronov1929}) que sa solution p\'{e}riodique est repr\'{e}sent\'{e}e dans le plan de phase par un \textit{cycle limite stable} ce qui constitue une condition n\'{e}cessaire \`{a} l'\'{e}tablissement d'un r\'{e}gime stable d'ondes entretenues.\\

Outre Atlantique, l'ing\'{e}nieur am\'{e}ricain Lee de Forest (1873-1961) d\'{e}pose le 15 janvier 1907 un brev\^{e}t n$^\circ$ 841386 intitul\'{e} ``Wireless Telegraphy'' pour un dispositif permettant l'amplification d'un signal \'{e}lectrique qu'il appelle \textit{audion}\footnote{L'audion g\'{e}n\'{e}rateur est le premier tube \'{e}lectronique de type triode.}. Lorsqu'\'{e}clate le premier conflit mondial, un mod\`{e}le d'\textit{audion} est confi\'{e}, dans des conditions rocambolesques\footnote{Voir \`{a} ce sujet la th\`{e}se de Ginoux \cite{Gith}.}, au G\'{e}n\'{e}ral Gustave Ferri\'{e} (1868-1932) alors directeur technique du service de la Radiot\'{e}l\'{e}graphie militaire qui demande au physicien Henri Abraham (1868-1943) de le reproduire et de le perfectionner. Ainsi modifi\'{e}, l'\textit{audion} va alors peu \`{a} peu s'imposer dans le corps des \og transmissions \fg{} sous le nom de \textit{lampe \`{a}  trois \'{e}lectrodes} puis de \textit{lampe T.M.\footnote{T\'{e}l\'{e}graphie Militaire.}}\\

Dans une note pr\'{e}sent\'{e}e le 14 avril 1919 lors d'une s\'{e}ance de l'Acad\'{e}mie des Sciences de Paris, Janet \cite{Janet1919} met en \'{e}vidence l'analogie\footnote{D\`{e}s 1891, Curie \cite{Curie} exposait d\'{e}j\`{a} les principes de l'analogie \'{e}lectrom\'{e}canique.} que pr\'{e}sentent la \textit{machine s\'{e}rie-dynamo}, l'\textit{arc chantant} et la \textit{lampe \`{a} trois \'{e}lectrodes}. Il en d\'{e}duit alors que ces trois dispositifs analogues, si\`{e}ge d'oscillations entretenues, sont r\'{e}gis par une seule et m\^{e}me \'{e}quation qu'il \'{e}tablit de façon incompl\`{e}te puisqu'il n'est alors pas en mesure d'expliciter la caract\'{e}ristique d'oscillation de la lampe \`{a} trois \'{e}lectrodes, ni celle de l'arc chantant.\\

En avril 1919, dans son article \og Sur les syst\`{e}mes \`{a} oscillations persistantes, et en particulier sur les oscillations entretenues par auto-amorçage \fg{}, Blondel \cite{Blondel1919a} introduit la terminologie \og auto-entretenue \fg{} et donne une seconde d\'{e}finition plus pr\'{e}cise du concept de \textit{syst\`{e}me auto-entretenu} qu'il illustre \`{a} partir d'un exemple m\'{e}taphorique : le siphon auto-amorceur (le vase de Tantale). Il analyse cette fois la stabilit\'{e} de l'amplitude des oscillations entretenues par un arc chantant puis par une lampe \`{a} trois \'{e}lectrodes en se plaçant dans le plan complexe et obtient en coordonn\'{e}es polaires des portions de spirales logarithmiques d\'{e}crivant \`{a} nouveau des cycles.\\

En novembre 1919, est publi\'{e}e aux \textit{Comptes Rendus} une note de Blondel \cite{Blondel1919c} intitul\'{e}e \og Amplitude du courant oscillant produit par les audions g\'{e}n\'{e}rateurs \fg{} dans laquelle il propose de mod\'{e}liser la caract\'{e}ristique d'oscillation de la lampe \`{a} trois \'{e}lectrodes \`{a} partir d'une s\'{e}rie \`{a} termes impairs (\textit{quintique}) ce qui lui permet d'\'{e}tablir, un an avant Balthazar Van der Pol, l'\'{e}quation diff\'{e}rentielle de ses oscillations.\\

En effet, dans un article intitul\'{e} ``A theory of the amplitude of free and forced triode vibrations'' achev\'{e} en juillet 1920 et publi\'{e} en novembre et d\'{e}cembre de la m\^{e}me ann\'{e}e, Van der Pol \cite{VdP1920} effectue la mise en \'{e}quation des oscillations entretenues par une triode en repr\'{e}sentant sa caract\'{e}ristique d'oscillation au moyen d'une fonction \textit{cubique}.\\

Le 28 juin 1920 paraissait en France l'ouvrage de Jean-Baptiste Pomey \cite{Pomey1920} intitul\'{e} : \og Introduction \`{a} la th\'{e}orie des courants t\'{e}l\'{e}phoniques et de la radiot\'{e}l\'{e}graphie \fg{}. Ancien \og \'{e}l\`{e}ve-ing\'{e}nieur \fg{} de l'\'{E}cole Sup\'{e}rieure Professionnelle des Postes et T\'{e}l\'{e}graphes (promotion 1883), Pomey \'{e}tait devenu professeur d'\'{e}lectricit\'{e} th\'{e}orique au sein m\^{e}me de cette \'{e}cole aux c\^{o}t\'{e}s d'Henri Poincar\'{e} avant d'en prendre la direction de 1924 \`{a} 1926. Dans le chapitre XIX intitul\'{e} \og G\'{e}n\'{e}ration des oscillations entretenues \fg{} Pomey \cite[p. 372]{Pomey1920} proposait pour repr\'{e}senter la caract\'{e}ristique de l'arc chantant de faire appel \`{a} une fonction cubique, pr\'{e}c\'{e}dant ainsi Van der Pol \cite{VdP1920} de quelques semaines.
\vspace{0.085in}

En septembre 1925, Pomey adressait un courrier au math\'{e}maticien \'{E}lie Cartan lui demandant d'\'{e}tudier les conditions d'entretien d'oscillations pour l'\'{e}quation diff\'{e}rentielle \'{e}tablie par Janet \cite{Janet1919} quelques ann\'{e}es auparavant. La r\'{e}ponse ne se fit pas attendre. Moins d'une semaine plus tard, Cartan faisait parvenir \`{a} Pomey le texte d'une \og Note sur la g\'{e}n\'{e}ration des oscillations entretenues \fg{} \cite{Cartan} \og r\'{e}dig\'{e}e\footnote{Au cours d'une entrevue accord\'{e}e \`{a} M. C. Gilain le samedi 30 septembre 2000, Henri Cartan aurait reconnu ne pas avoir particip\'{e} \`{a} la r\'{e}daction de cette note.}\fg{} avec son fils Henri et dans laquelle il \'{e}tablissait l'existence d'une solution p\'{e}riodique pour cette \'{e}quation\footnote{Des \'{e}changes \'{e}pistolaires entre Pomey et \'{E}lie Cartan retrouv\'{e}s lors de la disparition de son fils Henri en 2008 ont conduit \`{a} reconsid\'{e}rer les circonstances dans lesquelles E. Cartan s'\'{e}tait int\'{e}ress\'{e} \`{a} ce probl\`{e}me. Voir Ginoux \textit{et al.} \cite{GiWa}}.

\vspace{0.085in}

En mars 1926, reprenant la mod\'{e}lisation employ\'{e}e par Pomey \cite{Pomey1920}, Blondel \cite{Blondel1926} \'{e}tablit pour l'arc musical de Duddell \cite{Duddell1900a,Duddell1900b} une \'{e}quation diff\'{e}rentielle analogue \`{a} celle qu'obtient concomitamment Van der Pol \cite{VdP1926} pour les oscillations de la triode.

\newpage

En mars 1926, Van der Pol \cite{VdP1926} met en lumi\`{e}re le ph\'{e}nom\`{e}ne d'oscillations de relaxation dont il fournit l'\'{e}quation \og prototype \fg{} dans sa c\'{e}l\`{e}bre publication \'{e}ponyme (\textit{Philosphical Magazine, 1926}). Au cours des diff\'{e}rentes conf\'{e}rences qu'il effectue en France entre 1928 et 1930, il s'attache \`{a} \'{e}numerer les dispositifs susceptibles d'engendrer un tel ph\'{e}nom\`{e}ne et reconna\^{i}t que l'\textit{arc sifflant} d\'{e}couvert par Blondel en 1905 est bien le si\`{e}ge d'oscillations de relaxation puis choisit le vase de Tantale comme exemple m\'{e}taphorique. Van der Pol pr\'{e}cise ensuite le m\'{e}canisme d'entretien de ces oscillations en soulignant l'aspect non lin\'{e}aire que pr\'{e}sente la caract\'{e}ristique d'oscillation de ces dispositifs. N\'{e}anmoins, bien qu'il ait repr\'{e}sent\'{e} graphiquement la solution p\'{e}riodique de son \'{e}quation dans le plan de phase, il ne r\'{e}alise pas, jusqu'\`{a} la publication d'Andronov, qu'il s'agit d'un \textit{cycle limite} de Poincar\'{e}.\\

En mai 1928, Alfred Li\'{e}nard \cite{Liénard1928} d\'{e}montrait l'existence et l'unicit\'{e} de la solution p\'{e}riodique d'une \'{e}quation diff\'{e}rentielle de forme plus g\'{e}n\'{e}rale\footnote{Aujourd'hui connue sous le nom d'\'{e}quation de Li\'{e}nard-Van der Pol.} que celle de Van der Pol \cite{VdP1926} en prenant pour point de d\'{e}part la contribution de MM. Cartan \cite{Cartan}. Cependant, bien que la description que donnait Li\'{e}nard \cite[p. 906]{Liénard1928} de la solution p\'{e}riodique corresponde exactement \`{a} la d\'{e}finition d'un \textit{cycle limite} de Poincar\'{e} il ne faisait absolument pas usage de cette expression.\\

En octobre 1929, dans une note\footnote{Ce r\'{e}sultat avait d\'{e}j\`{a} \'{e}t\'{e} pr\'{e}sent\'{e} par Andronov \cite{Andronov1928} quelques mois auparavant.} intitul\'{e}e \og Les cycles limites de Poincar\'{e} et la th\'{e}orie des oscillations auto-entretenues \fg{} pr\'{e}sent\'{e}e devant les membres de l'Acad\'{e}mie des Sciences, le math\'{e}maticien russe Aleksandr' Andronov \cite{Andronov1929} \'{e}tablissait (vingt ans apr\`{e}s Poincar\'{e} \cite{P6}) que la solution p\'{e}riodique d'un oscillateur auto-entretenu correspond dans le plan de phase \`{a} un \textit{cycle limite stable} de Poincar\'{e}.\\

Dans une lettre r\'{e}cemment d\'{e}couverte, dat\'{e}e d'Avril 1931 et adress\'{e}e par Andr\'{e} Blondel \`{a} \'{E}lie Cartan il sugg\'{e}rait qu'il avait entraperçu les oscillations de relaxation en mettant en \'{e}vidence un nouveau type discontinu d'oscillations pour l'\textit{arc sifflant}.\\

L'objet de cet article est donc d'analyser les principaux travaux de Blondel dans le domaine des oscillations non lin\'{e}aires afin de mesurer leur importance voire leur influence sur la conceptualisation des oscillations auto-entretenues et des oscillations de relaxation.

\newpage

\section{Sur le ph\'{e}nom\`{e}ne de l'arc chantant}

En 1883, alors \^{a}g\'{e} de vingt ans, Andr\'{e} Blondel (1861-1938) int\`{e}gre l'\'{E}cole Polytechnique, apr\`{e}s avoir
\'{e}t\'{e} reçu la m\^{e}me ann\'{e}e \`{a} l'\'{E}cole Normale sup\'{e}rieure. Il entre ensuite \`{a} l'\'{E}cole des Ponts et
Chauss\'{e}es en 1885, d'o\`{u} il sort major en 1888. Licenci\'{e} \`{e}s Sciences Math\'{e}matiques en 1885 et \`{e}s Sciences Physiques en 1889 il suit les cours d'Henri Poincar\'{e} de 1888 \`{a} 1889 avant de commencer sa carri\`{e}re d'ing\'{e}nieur au Service central des Phares et Balises, un poste qui relevait alors de la Direction G\'{e}n\'{e}rale des Ponts et Chauss\'{e}es. Cependant, la vie et la carri\`{e}re de Blondel ont \'{e}t\'{e} marqu\'{e}es par un fait frappant comme le rappellent ses principaux biographes \cite{Bethenod1938,Broglie}. En effet, d\`{e}s 1892 il semble avoir \'{e}t\'{e} atteint d'une maladie myst\'{e}rieuse se manifestant par une semi-paralysie des jambes qui aurait \'{e}t\'{e} la cons\'{e}quence d'une chute de cheval et qui l'aurait contraint \`{a} rester presque alit\'{e} durant vingt-sept ann\'{e}es.

\begin{quote}

\og
Il fut au Service des Phares et Balises de 1889 \`{a} 1927, mais il n'est presque jamais venu \`{a} son bureau et n'a presque jamais fait de tourn\'{e}es en mer ; cependant les premiers radiophares lui sont dus. Il fut Professeur \`{a} l'\'{E}cole des Ponts et Chauss\'{e}es de 1893 \`{a} 1929, mais il n'y a jamais profess\'{e} : cependant tous les cours d'\'{E}lectricit\'{e} dependent de ses expos\'{e}s. Il fit les recherches exp\'{e}rimentales les plus profondes et a mis au point les appareils les plus d\'{e}licats : cependant il n'alla jamais \`{a} son laboratoire. \fg{} \cite[p. 4]{Broglie}

\end{quote}

Ainsi attach\'{e} au corps des Ponts et Chauss\'{e}es au service des Phares et Balises, il oriente ses recherches vers l'\'{e}tude de l'arc \'{e}lectrique \og  en vue d'une application sp\'{e}ciale d'\'{e}clairage  \fg{}  \cite[p. 552]{Blondel1891}. De 1891 \`{a} 1905, Blondel \cite{Blondel1891,Blondel1893b,Blondel1897,Blondel1905a,Blondel1905b,Blondel1905c} publie toute une s\'{e}rie de travaux sur l'arc \`{a} courant alternatif puis sur l'arc \`{a} courant continu. Le 7 juillet 1905, Blondel pr\'{e}sente devant les membres de la Soci\'{e}t\'{e} française de Physique\footnote{Cet expos\'{e} avait \'{e}t\'{e} pr\'{e}c\'{e}d\'{e} d'une courte note pr\'{e}sent\'{e}e le 13 juin 1905 par \'{E}leuth\`{e}re Mascart (1837-1908) devant les membres de l'Acad\'{e}mie des Sciences. Voir Blondel \cite{Blondel1905a}. Une version compl\`{e}te de ce travail avait alors \'{e}t\'{e} publi\'{e}e dans la \textit{L'\'{E}clairage \'{E}lectrique}. Voir Blondel \cite{Blondel1905c}.} un imposant m\'{e}moire \og Sur les ph\'{e}nom\`{e}nes de l'arc chantant \fg{} \cite{Blondel1905c} dans lequel il met en \'{e}vidence deux types extr\^{e}mes d'arcs, l'un continu correspondant \`{a} l'arc chantant de Duddell \cite{Duddell1900a,Duddell1900a,Duddell1901} et caract\'{e}ris\'{e} par des oscillations \og presque sinusoïdales \fg{} \cite{Blondel1905b}, l'autre discontinu qu'il qualifie d'\textit{arc sifflant} et dont la p\'{e}riode ne se conforme plus \`{a} la formule de Thomson \cite[p. 393]{Thomson} ni \`{a} celle de Duddell \cite{Duddell1901}.

\newpage

Pour y parvenir, Blondel fait appel d'une part \`{a} l'analyse oscillographique (voir Fig. 2) et, introduit d'autre part la notion de \textit{caract\'{e}ristique dynamique} (voir Fig. 5) permettant d'\'{e}tudier la stabilit\'{e} des r\'{e}gimes d'oscillations de l'arc.\\

Pour \'{e}tudier ce qu'il consid\`{e}re comme la \og nature intime du ph\'{e}nom\`{e}ne \fg{}\\ Blondel \cite[p. 44]{Blondel1905c} utilise le montage comportant un  arc $H$, un condensateur de capacit\'{e} $C$, un rh\'{e}ostat $R$, deux bobines d'induction $L$ et $l$ et o\`{u} $ABDF$ repr\'{e}sente le circuit d'alimentation provenant d'un secteur \`{a} courant continu et $BCD$ le circuit d'oscillation (voir Fig. 1.).

\begin{figure}[htbp]
\centerline{\includegraphics[width=12.59cm,height=4.23cm]{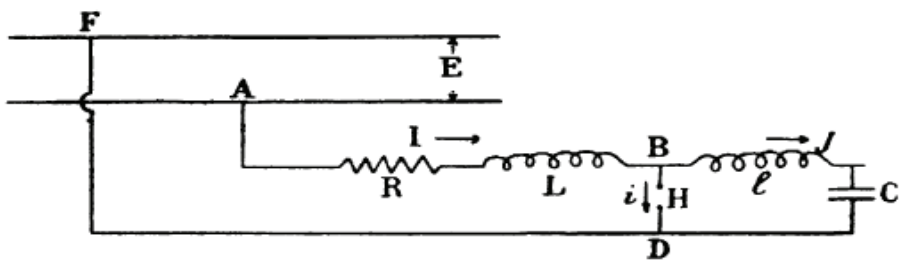}}
\caption[Montage des exp\'{e}riences sur l'arc chantant]{Montage des exp\'{e}riences sur l'arc chantant,\\ \centering{d'apr\`{e}s Blondel \cite[p. 44]{Blondel1905c}.}}
\label{fig1}
\end{figure}

En reliant \`{a} ce dispositif un oscillographe triple bifilaire qu'il a fait construire tout sp\'{e}cialement il obtient les clich\'{e}s oscillographiques de la tension $u$ aux bornes de l'arc et de l'intensit\'{e} du courant $i$ qui le traverse ainsi que l'intensit\'{e} du courant $j$ dans le condensateur $C$. Ceci lui permet d'\'{e}tablir l'existence de deux r\'{e}gimes d'oscillation.\\

Le premier type continu : \textit{arc musical}, (Fig. 2) \og auquel correspond un son soutenu assez pur, et qui est, \`{a} proprement parler, l'arc \textit{musical} de Duddell, donne lieu \`{a} des courbes de courant dans l'arc et le condensateur pr\'{e}sentant des formes \textit{continues}, presque sinusoïdales, sans que l'intensit\'{e} de l'arc descende \`{a} z\'{e}ro ou tout au moins reste nulle pendant un temps appr\'{e}ciable ; les variations de la tension aux bornes de l'arc sont contenues entre des limites tr\`{e}s rapproch\'{e}es. \fg{} \cite[p. 79]{Blondel1905b}.

\begin{figure}[htbp]
\centerline{\includegraphics[width=11cm,height=13.3cm]{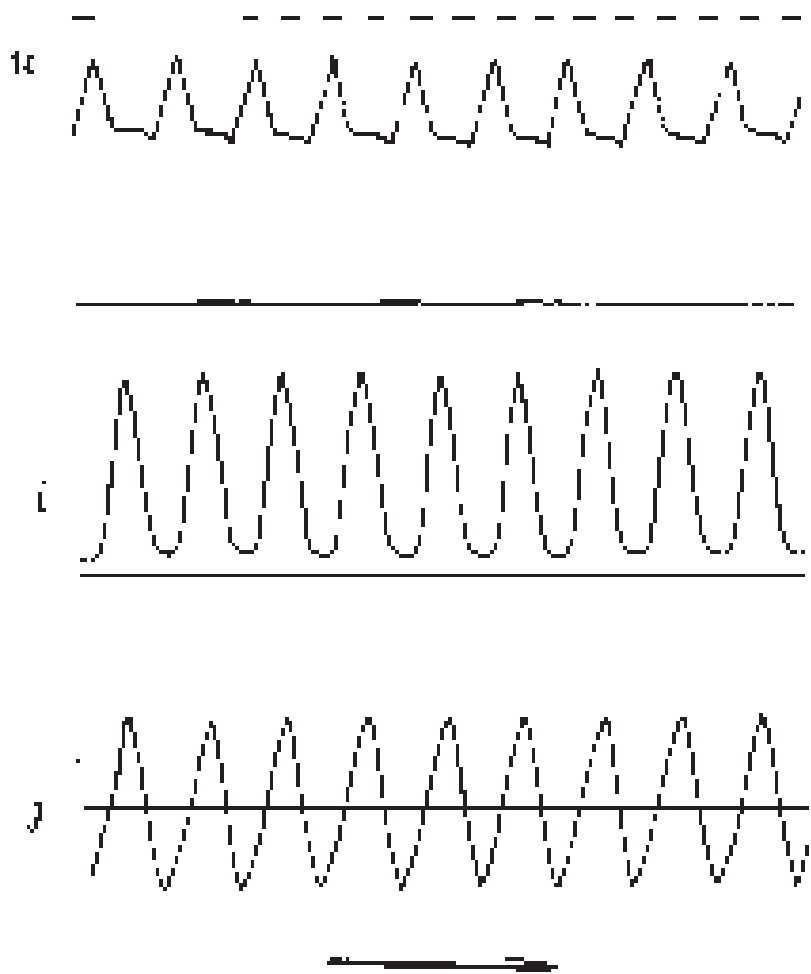}}
\caption[Premier type~: arc musical]{Courbes du premier type~: \textit{arc musical},\\ \centering{d'apr\`{e}s Blondel \cite[p. 79]{Blondel1905b}}.}
\label{fig2}
\end{figure}

\newpage

Blondel en d\'{e}duit que bien que la p\'{e}riode de l'arc chantant soit essentiellement variable et mal d\'{e}finie, elle peut, dans le cas du ph\'{e}nom\`{e}ne continu (premier cas) \^{e}tre pr\'{e}vue approximativement par la formule de Duddell\footnote{Duddell avait modifi\'{e} la formule de Thomson \cite[p. 393]{Thomson} : $T = 2\pi \sqrt{LC}$ afin de tenir compte de la r\'{e}sistance du circuit.} \cite{Duddell1901} :

\[
T =  \dfrac{2\pi}{\sqrt{\dfrac{1}{LC} - \dfrac{R^2}{4L^2}}}
\]

\newpage

Le second type discontinu : \textit{arc sifflant}\footnote{Par la suite, Blondel lui donnera \'{e}galement le nom d'\textit{arc gr\'{e}sillant}.}, (Fig. 3) \og auquel correspond un son plus strident ou sifflant, est un ph\'{e}nom\`{e}ne discontinu caract\'{e}ris\'{e} par ce que le courant de l'arc $i$ pr\'{e}sente des points anguleux et des z\'{e}ros de temps appr\'{e}ciable, pendant lesquels le~courant de charge $j$ pr\'{e}sente ordinairement des m\'{e}plats, tandis que la tension entre les \'{e}lectrodes $u$ subit une double oscillation de grande amplitude allant souvent au-dessous de z\'{e}ro ou au-dessus de la force \'{e}lectromotrice
d'alimentation. \fg{} \cite[p. 80]{Blondel1905b}

\begin{figure}[htbp]
\centerline{\includegraphics[width=11cm,height=13.84cm]{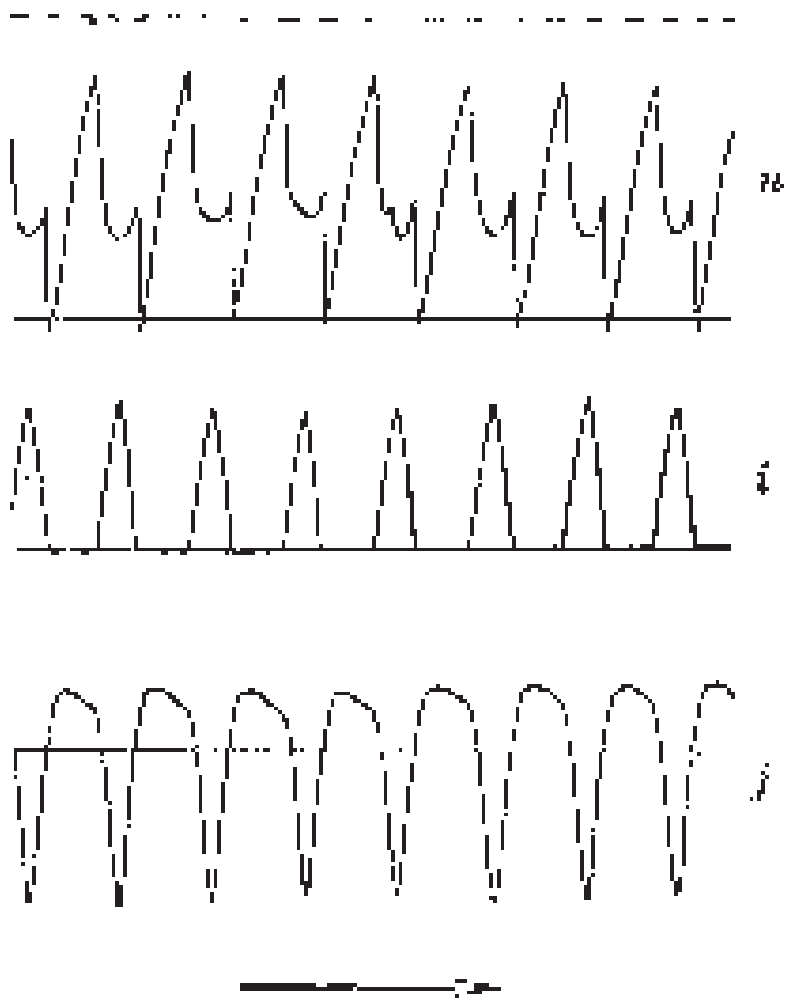}}
\caption[Second type~: arc sifflant]{Courbes du second type~: \textit{arc sifflant},\\ \centering{d'apr\`{e}s Blondel \cite[p. 78]{Blondel1905b}}.}
\label{fig3}
\end{figure}

Puis, il ajoute que :

\begin{quote}

\og Pendant l'extinction (ABC), la tension aux bornes de l'arc tombe brusquement aux environs de z\'{e}ro, puis s'\'{e}l\`{e}ve plus lentement \`{a} une valeur tr\`{e}s notablement sup\'{e}rieure \`{a} sa valeur de r\'{e}gime ; au moment de l'allumage (CD) elle tombe brusquement \`{a} cette valeur. \fg{}\\ \vphantom{} \hfill \cite[p. 48]{Blondel1905c}

\end{quote}

\begin{figure}[htbp]
\centerline{\includegraphics[width=5.392cm,height=6.72cm]{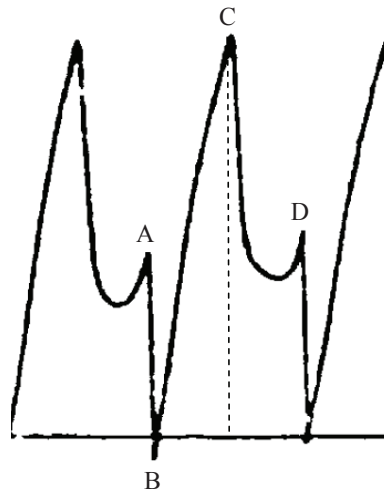}}
\caption[Extinction et allumage de l'arc sifflant]{Extinction et allumage de l'arc sifflant,\\ \centering{d'apr\`{e}s Blondel \cite[p. 78]{Blondel1905b}}.}
\label{fig4}
\end{figure}

Blondel \'{e}tablit ensuite une analogie entre ces oscillations du second type produites par l'\textit{arc sifflant} et le ph\'{e}nom\`{e}ne de la \textit{d\'{e}charge fractionn\'{e}e}\footnote{Voir Gaugain \cite{Gaugain}.} mis en \'{e}vidence en 1866 par le physicien français Jean-Moth\'{e}e Gaugain (1811-1880). Il explique alors que la p\'{e}riodicit\'{e} de ces oscillations n'a aucun rapport direct avec les constantes du circuit, contrairement \`{a} la formule de Duddell, mais d\'{e}pend principalement de la composition du circuit d'alimentation et conclut son article par cette phrase :

\begin{quote}
\og Les constantes du circuit d'alimentation d\'{e}terminent les intervalles entre les groupes d'oscillations. \fg{}\\
\vphantom{} \hfill \cite[p. 97]{Blondel1905b}

\end{quote}

Blondel s'int\'{e}resse \'{e}galement au passage du r\'{e}gime du second type (discontinu) au premier (continu) et inversement, c'est-\`{a}-dire, des oscillations non-sinusoïdales aux oscillations sinusoïdales.

Il remarque alors que dans le cas discontinu la courbe de tension $u$ aux bornes de l'arc n'est pas sym\'{e}trique comme celle du courant qui le traverse et reste plus \'{e}lev\'{e}e pendant la phase montante du courant que pendant la phase descendante. Il d\'{e}duit de ceci d'une part que la \textit{r\'{e}sistance} de l'arc n'est pas constante mais variable. Ce fait avait \'{e}t\'{e} \'{e}tabli par Luggin \cite[p. 568]{Luggin} quelques ann\'{e}es auparavant qui avait introduit le concept de \textit{r\'{e}sistance n\'{e}gative} et avait ensuite \'{e}t\'{e} corrobor\'{e} par Blondel \cite[p. 515]{Blondel1897}. D'autre part, que le cycle de r\'{e}gimes (ABCD) pendant une p\'{e}riode n'est pas r\'{e}versible. Il \'{e}crit :

\begin{quote}

\og On pourrait d'ailleurs tracer ce cycle au moyen des deux courbes de l'intensit\'{e} $i$ et de la tension $u$ en reportant en abscisses les valeurs de $i$ et en ordonn\'{e}es les valeurs $u$ correspondantes. \fg{} \cite[p. 48]{Blondel1905c}

\end{quote}

C'est ainsi qu'il introduit la notion de \textit{caract\'{e}ristique dynamique} (voir Fig. 5) qui consiste \`{a} se placer dans le plan de phase de Poincar\'{e} \cite[p. 168]{P4} pour expliquer les ph\'{e}nom\`{e}nes observ\'{e}s.

\vspace{0.1in}

\begin{figure}[htbp]
\centerline{\includegraphics[width=14.355cm,height=8.973cm]{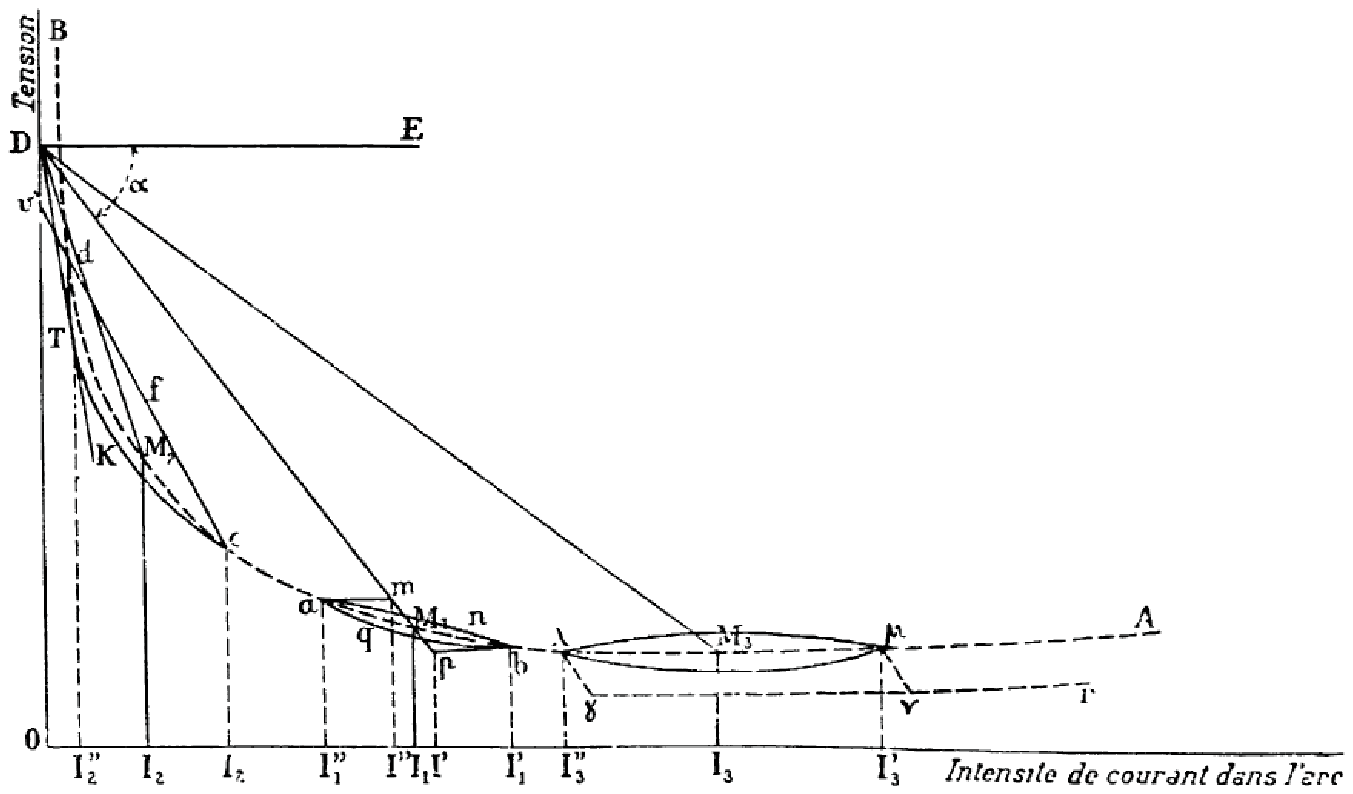}}
\caption[Caract\'{e}ristique dynamique de l'arc]{Caract\'{e}ristique dynamique de l'arc, \\ \centering{d'apr\`{e}s Blondel \cite[p. 53]{Blondel1905c}.}}
\label{fig5}
\end{figure}

\newpage

Il constate \`{a} l'oscillographe (comme le confirmera plus tard Pomey \cite[p. 384]{Pomey1920}) que lorsqu'on fait varier l'intensit\'{e} m\^{e}me tr\`{e}s faiblement entre deux points limites $I'_1 $ et $I''_1 $, le point r\'{e}gime $M_1$ d\'{e}crit, non pas une droite, mais un petit cycle $anbqa$ parcouru dans ce sens\footnote{Il est important de remarquer que c'est le m\^{e}me sens de parcours r\'{e}trograde qu'empruntera la solution p\'{e}riodique de l'\'{e}quation de Van der Pol \cite{VdP1926}.}. Il \'{e}crit :

\begin{quote}

\og Le graphique montre aussi qu'une partie du courant d'alimentation sert \`{a} compenser les pertes d'\'{e}nergie par effet Joule ou autres dans le circuit oscillant, gr\^{a}ce au fait que la branche de charge \textit{amb} du cycle est au dessus de la branche de d\'{e}charge \textit{bqa} ; le condensateur reçoit ainsi plus d'\'{e}nergie qu'il n'a \`{a} en restituer.
\fg{} \cite[p. 54]{Blondel1905c}

\end{quote}

Cette description d'un dispositif dans lequel une partie de l'\'{e}nergie produite est utilis\'{e}e pour compenser les pertes et entretenir ainsi les oscillations constitue les pr\'{e}mices de la d\'{e}finition que Blondel \cite[p. 120]{Blondel1919c} donnera ensuite d'un syst\`{e}me \textit{auto-entretenu}. Enfin, en ce qui concerne l'amplitude et la fr\'{e}quence des oscillations, il tire la conclusion suivante~:

\begin{quote}

\og
L'amplitude de l'oscillation \textit{ab} s'\'{e}tablit elle-m\^{e}me de façon que l'aire de la boucle \'{e}gale l'\'{e}nergie 					 perdue dans le circuit d'oscillation, et la fr\'{e}quence est fonction surtout de la vitesse de charge et de 				 d\'{e}charge du condensateur, d\'{e}termin\'{e}e par les constantes du circuit oscillant, mais modifi\'{e}e certainement dans une certaine mesure par la composition du circuit d'alimentation et par la forme du cycle \textit{ambn} d\'{e}crit par le r\'{e}gime de l'arc. \fg{} \\ \vphantom{} \hfill \cite[p. 54]{Blondel1905c}

\end{quote}

Durant la guerre de 1914-1918, lorsque les services de la Radiot\'{e}l\'{e}graphie militaire dirig\'{e} par son camarade d'\'{e}cole, le G\'{e}n\'{e}ral Ferri\'{e} (alors Colonel) envisagent de substituer la lampe \`{a} trois \'{e}lectrodes \`{a} l'arc chantant dans les postes de transmission, Blondel entreprend l'\'{e}tude de ce nouveau dispositif \og par analogie avec la th\'{e}orie d\'{e}j\`{a} connue de l'arc chantant \fg{} \cite[p. 676]{Blondel1919b} transposant ainsi la plupart des r\'{e}sultats qu'il avait obtenus.

\newpage

\section{Le multivibrateur d'Abraham et Bloch}

Au d\'{e}but du premier conflit mondial, le physicien Henri Abraham est envoy\'{e} \`{a} Lyon par le Colonel Gustave Ferri\'{e}, alors directeur technique du service de Radiot\'{e}l\'{e}graphie militaire, avec pour mission de reproduire et de perfectionner l'audion de Lee de Forest dans le but de r\'{e}aliser une production \`{a} grande \'{e}chelle permettant d'\'{e}quiper rapidement les r\'{e}giments de transmissions d'un dispositif plus fiable et plus efficace. D\`{e}s le mois de f\'{e}vrier 1915, Abraham s'\'{e}tait acquit\'{e} de cette t\^{a}che en \'{e}laborant avec François P\'{e}ri (1870-1938) et Jacques Biguet (1880-1970) la lampe T.M. \'{e}galement appel\'{e}e \og lampe française \fg{} qui atteignit pour l'\'{e}poque un tel degr\'{e} de perfection et de fiabilit\'{e} qu'elle fut adopt\'{e}e par l'arm\'{e}e française puis par les arm\'{e}es alli\'{e}es et fut produite en France \`{a} plus d'un million d'exemplaires pendant la dur\'{e}e du conflit. De retour \`{a} Paris en mai 1915, Abraham reprend ses fonctions de directeur du Laboratoire de Physique de l'\'{E}cole Normale Sup\'{e}rieure et, poursuivant ses recherches, invente\footnote{Bien que les publications d'Abraham et Bloch \cite{AbrahamBloch} datent de 1919, ce dispositif a \'{e}t\'{e} \'{e}labor\'{e} entre novembre et d\'{e}cembre 1917 comme en attestent les diff\'{e}rents documents class\'{e}s \og secret d\'{e}fense \fg{} r\'{e}cemment retrouv\'{e}s. Voir Ginoux \cite{Gith}.} avec son coll\`{e}gue Eug\`{e}ne Bloch (1878-1944) le \textit{multivibrateur astable}. Cet appareil est constitu\'{e} de deux lampes T.M. dont chaque grille est reli\'{e}e \`{a} la plaque de l'autre par un condensateur. Un tel montage produit des oscillations tr\`{e}s riches en harmoniques. Abraham explique alors que c'est la raison pour laquelle il lui a donn\'{e} le nom de \textit{multivibrateur} (voir Fig. 6).\\

\begin{figure}[htbp]
\centerline{\includegraphics[width=8.704cm,height=7.216cm]{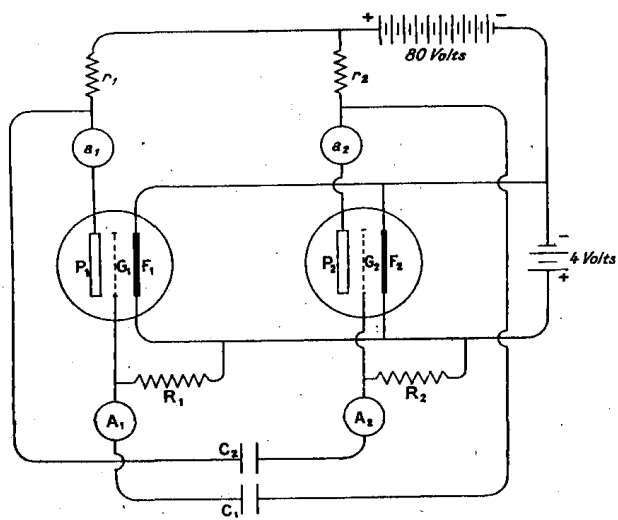}}
\caption[Multivibrateur d'Abraham et Bloch]{Multivibrateur, d'apr\`{e}s Abraham et Bloch \cite[p. 254]{AbrahamBloch}.}
\label{fig6}
\end{figure}

\`{A} cette \'{e}poque on consid\'{e}rait que la p\'{e}riode d'oscillations d'un circuit comportant un condensateur de capacit\'{e} $C$ et une bobine d'inductance $L$ \'{e}tait donn\'{e}e par la formule de Thomson ou de Duddell. Cependant, Abraham et Bloch d\'{e}montrent que les oscillations produites par le multivibrateur poss\`{e}dent une p\'{e}riode qui ne se conforme pas \`{a} ces formules. Concernant les inversions des courants de plaques $P_{1}$ et $P_{2}$ reproduites sur la Fig. 7. ils \'{e}crivent :

\begin{quote}

\og
On constate une s\'{e}rie d'inversions tr\`{e}s brusques des courants s\'{e}par\'{e}es par des longs intervalles pendant lesquelles la variation d'intensit\'{e} du courant est tr\`{e}s lente\footnote{Rapport E.C.M.R. n\r{} 2949, novembre 1917. Voir Ginoux \cite{Gith}.}.
\fg{}

\end{quote}

Ils ajoutent que les intervalles de temps qui s\'{e}parent ces inversions correspondent aux dur\'{e}es de charges et de d\'{e}charges des condensateurs $C_{1}$ et $C_{2}$ \`{a} travers les r\'{e}sistances $R_{1}$ et $R_{2}$. Ce qui semble faire \'{e}cho \`{a} ce qu'avait \'{e}crit Blondel \cite[p. 97]{Blondel1905b} (voir p. 11).

\begin{figure}[htbp]
\centerline{\includegraphics[width=13.27cm,height=4cm]{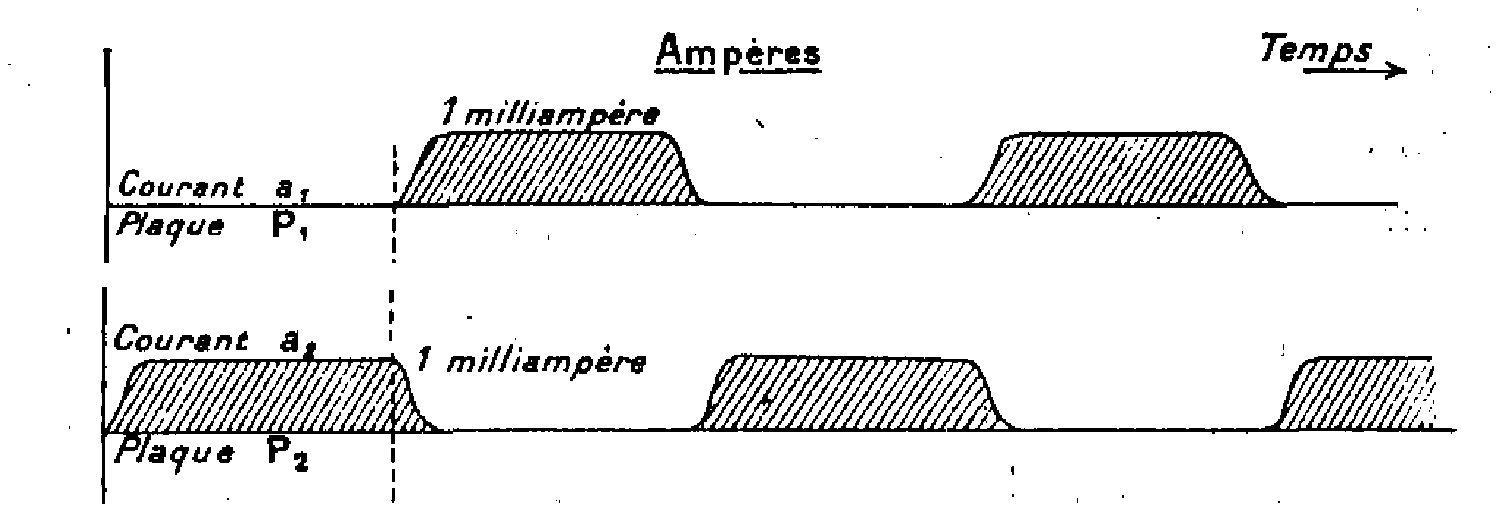}}
\caption[Inversions des courants de plaques P$_{1}$ et P$_{2}$]{Inversions des courants de plaques P$_{1}$ et P$_{2}$,
\\ \centering{d'apr\`{e}s Abraham et Bloch \cite[p. 256]{AbrahamBloch}.}}
\label{fig7}
\end{figure}

\newpage

Ils en concluent\footnote{Rapport E.C.M.R. n\r{} 2949, novembre 1917.} :\\

\begin{center}

\og La p\'{e}riode du syst\`{e}me est donc de l'ordre de $C_1 R_1 + C_2 R_2$. \fg{}

\end{center}

\vspace{0.1in}

Au sortir du premier conflit mondial, l'ing\'{e}nieur français Paul Janet va mettre en \'{e}vidence une analogie \'{e}lectrotechnique entre trois dispositifs : la machine s\'{e}rie-dynamo, l'arc chantant et la lampe \`{a} trois \'{e}lectrodes permettant d'une part de transposer tous les r\'{e}sultats obtenus pr\'{e}c\'{e}demment pour l'arc chantant \`{a} la lampe \`{a} trois \'{e}lectrodes et, d'autre part de d\'{e}crire des ph\'{e}nom\`{e}nes en apparence tr\`{e}s diff\'{e}rents par une unique \'{e}quation diff\'{e}rentielle.

\section{Sur une analogie \'{e}lectrotechnique} \hfill

Normalien \`{a} vingt-deux ans, agr\'{e}g\'{e} en physique, Paul Janet (1863-1937), fils du philosophe Paul Janet (1823-1899), soutient en 1890 sa th\`{e}se de doctorat intitul\'{e}e~: \og \'{E}tude th\'{e}orique et exp\'{e}rimentale sur l'aimantation transversale des conducteurs magn\'{e}tiques \fg{}, \`{a} la facult\'{e} des Sciences de Paris. Il est alors nomm\'{e} ma\^{i}tre de conf\'{e}rences \`{a} Grenoble o\`{u} il inaugure en 1892, contre l'avis du doyen François Raoult, le tout premier cours d'\'{E}lectricit\'{e} Industrielle. Le succ\`{e}s est tel que François Raoult est amen\'{e} \`{a} demander officiellement la cr\'{e}ation des cours et d'un laboratoire d'\'{e}lectricit\'{e} d\`{e}s la rentr\'{e}e suivante. Deux ans plus tard, Paul Janet devient directeur de la toute nouvelle \'{E}cole Sup\'{e}rieure d'\'{E}lectricit\'{e}\footnote{Il occupera ce poste jusqu'en 1937, date de son d\'{e}c\`{e}s.} \`{a} Paris o\`{u} il poursuit son enseignement. D\`{e}s 1894, il s'int\'{e}resse aux oscillations d'une \textit{machine s\'{e}rie-dynamo} qui est constitu\'{e}e par l'association d'une \textit{machine dynamo-\'{e}lectrique} jouant le r\^{o}le de g\'{e}n\'{e}rateur avec une \textit{machine magn\'{e}to-\'{e}lectrique} se comportant comme un moteur. Le 8 d\'{e}cembre 1919, il est \'{e}lu acad\'{e}micien libre. Quelques mois auparavant, le 14 avril, il pr\'{e}sente une note intitul\'{e}e : \og Sur une analogie \'{e}lectrotechnique des oscillations entretenues \fg{} \cite{Janet1919} dans laquelle il d\'{e}crit ainsi un ph\'{e}nom\`{e}ne qui avait \'{e}t\'{e} observ\'{e} un demi-si\`{e}cle auparavant par Jean-Marie Anatole G\'{e}rard-Lescuyer \cite{Lescuyer} :

\begin{quote}
\og Si l'on alimente, au moyen d'une g\'{e}n\'{e}ratrice excit\'{e}e en s\'{e}rie, un moteur \`{a} excitation s\'{e}par\'{e}e tournant \`{a} vide, on voit le moteur rapidement dans un sens, s'arr\^{e}ter, repartir en sens inverse, etc. \fg{} \cite[p. 764]{Janet1919}
\end{quote}

En r\'{e}alit\'{e}, quelques ann\'{e}es plus tard, Van der Pol \cite{VdP1928} consid\`{e}rera l'exp\'{e}rience expos\'{e}e par Janet comme un exemple d'\textit{oscillations de relaxation}. Cette note de Janet \cite{Janet1919} rev\^{e}t une importance consid\'{e}rable. En tout premier lieu dans la mesure o\`{u} elle met en lumi\`{e}re une \og analogie \'{e}lectrotechnique \fg{} entre les oscillations entretenues produites par une machine s\'{e}rie-dynamo et celles que pr\'{e}sentent un arc chantant ou une lampe \`{a} trois \'{e}lectrodes.

\begin{quote}

\og Il m'a sembl\'{e} int\'{e}ressant, \'{e}crit-il, de signaler les
analogies inattendues que pr\'{e}sente cette exp\'{e}rience avec les oscillations
entretenues si largement utilis\'{e}es aujourd'hui en t\'{e}l\'{e}graphie
sans fil, par exemple avec celles qui se produisent dans l'arc de Duddell ou
dans les lampes \`{a} trois \'{e}lectrodes employ\'{e}es comme
oscillateurs. \fg{}\\ \vphantom{} \hfill \cite[p. 764]{Janet1919}

\end{quote}

En second lieu, parce que Janet y souligne le transfert de technologie consistant \`{a} remplacer un composant \'{e}lectrom\'{e}canique (arc chantant) par ce que l'on appellera plus tard un tube \'{e}lectronique, ce qui constitue une v\'{e}ritable r\'{e}volution car l'arc chantant, du fait m\^{e}me de sa structure, rendait l'exp\'{e}rimentation complexe et d\'{e}licate et sa reproductibilit\'{e} difficile, voire impossible. Pour justifier cette \og analogie \'{e}lectrotechnique \fg{} il fonde son raisonnement sur une analogie plus ancienne qui concerne les composants des circuits~:

\begin{quote}

\og La production et l'entretien des oscillations dans tous ces syst\`{e}mes
tiennent essentiellement \`{a} la pr\'{e}sence, dans le circuit oscillant de
quelque chose d'analogue \`{a} une r\'{e}sistance n\'{e}gative~: or, la
dynamo-s\'{e}rie g\'{e}n\'{e}ratrice se comporte comme une r\'{e}sistance
n\'{e}gative, et, d'autre part, le moteur \`{a} excitation s\'{e}par\'{e}e
se comporte comme un condensateur.~Il est curieux de constater que ces deux
analogies ont \'{e}t\'{e} signal\'{e}es il y a longtemps, la premi\`{e}re
par M. P. Boucherot et la seconde par M. Maurice Leblanc.  \fg{} \cite[p. 764]{Janet1919}

\end{quote}

Il consid\`{e}re que pour qu'il y ait analogie dans les effets, $i.e.$ pour que l'on observe le m\^{e}me type d'oscillations dans la machine s\'{e}rie-dynamo, la triode et l'arc chantant il faut qu'il y ait analogie dans les causes. Or, puisque la~dynamo-s\'{e}rie se comporte comme une r\'{e}sistance n\'{e}gative, qui est la cause responsable des oscillations,
il y a bien analogie. Par cons\'{e}quent, \`{a} ces diff\'{e}rents dispositifs doit correspondre une seule et m\^{e}me \'{e}quation.

\begin{quote}

\og La mise en \'{e}quation du probl\`{e}me, dans le cas du syst\`{e}me
mat\'{e}riel indiqu\'{e} plus haut, est facile. Soit, $e=f\left( i \right)$
la force \'{e}lectromotrice de la dynamo-s\'{e}rie, R et L la r\'{e}sistance
et la self-induction du circuit, $\omega $ la vitesse angulaire du moteur
\`{a} excitation s\'{e}par\'{e}e~; on a \'{e}videmment

\[
Ri+L\frac{di}{dt}=e-k\omega
\]

\[
ki=K\frac{d\omega }{dt}
\]

d'o\`{u}

\begin{center}
\hfill
$L\dfrac{d^2i}{dt^2} + \left[ {R-{f}'\left( i \right)} \right]\dfrac{di}{dt}+\dfrac{k^2}{K}i=0$ \hfill (J$_{1}$).\fg
\end{center}
\vphantom{} \hfill  \cite[p. 765]{Janet1919}

\end{quote}

La premi\`{e}re \'{e}quation~que l'on peut r\'{e}\'{e}crire ainsi : $e=L\frac{di}{dt}+Ri+k\omega $ transcrit le fait que
pour expliquer compl\`{e}tement le ph\'{e}nom\`{e}ne il faut tenir compte
de~:\\

\begin{itemize}

\item[a.] la \textit{f.\'{e}.m.} de la dynamo~: $e=f\left( i \right)$~ ;\\

\item[b.] la \textit{f.c.\'{e}.m.} du moteur~: $Ri+k\omega $~ ;\\

\item[c.] la \textit{f.\'{e}.m.} de la bobine d'induction~: $L\dfrac{di}{dt}$.\\

\end{itemize}

Janet obtient ainsi une \'{e}quation (J$_{1}$) analogue \`{a} celle qu'avait \'{e}tablie Poincar\'{e} \cite{P6} vingt ans auparavant pour les oscillations entretenues par un arc chantant. On peut donc s'\'{e}tonner qu'il ne fasse pas r\'{e}f\'{e}rence \`{a} ce travail de Poincar\'{e} \cite{P6} alors qu'il cite ceux, plus anciens, de Leblanc (1899) et Boucherot (1904). En effet, lors d'une s\'{e}rie de Conf\'{e}rences (r\'{e}cemment red\'{e}couverte\footnote{Voir Ginoux \cite{Gith}.}) effectu\'{e}es en mai-juin 1908 \`{a} l'\'{E}cole Sup\'{e}rieure des Postes et T\'{e}l\'{e}graphes, Poincar\'{e} \cite[p. 390]{P6} avait \'{e}tabli l'\'{e}quation suivante pour d\'{e}crire les oscillations d'un arc chantant :

\vspace{0.1in}

\begin{center}

\hfill $L{x}''+\rho {x}'+\theta \left( {x}' \right) + Hx=0$ \hfill (HP$_{1}$)

\end{center}

\vspace{0.1in}

o\`{u} $L$ et $1/H$ correspondent respectivement \`{a} la self de la bobine d'induction de r\'{e}sistance interne $\rho$ et \`{a} la capacit\'{e} du condensateur plac\'{e}s en parall\`{e}le avec l'arc chantant. La variable $x$ repr\'{e}sente la charge du condensateur ($x'$ l'intensit\'{e} du courant dans la branche comportant le condensateur). Cette \'{e}quation diff\'{e}rentielle ($P_{1}$) est n\'{e}anmoins incompl\`{e}te puisque Poincar\'{e} n'a pas cherch\'{e} \`{a} expliciter la fonction $\theta(x')$, c'est-\`{a}-dire, la caract\'{e}ristique d'oscillation (\textit{f.\'{e}.m.}) de l'arc chantant.\\ Il explique alors :

\begin{quote}

\og On peut construire les courbes qui satisfont \`{a} cette \'{e}quation
diff\'{e}rentielle, \`{a} condition de conna\^{i}tre la fonction $\theta $.
Les oscillations entretenues correspondent aux courbes ferm\'{e}es, s'il y
en a. Mais toute courbe ferm\'{e}e ne convient pas, elle doit remplir
certaines conditions de stabilit\'{e} que nous allons \'{e}tudier. \fg{} \cite[p. 390]{P6}

\end{quote}

Puis, il d\'{e}montre que l'\'{e}tablissement d'un r\'{e}gime d'ondes entretenues est conditionn\'{e} par la pr\'{e}sence dans l'espace des phases d'une courbe ferm\'{e}e qui correspond exactement \`{a} la d\'{e}finition qu'il a donn\'{e}e quelques ann\'{e}es auparavant d'un \textit{cycle limite}\footnote{Voir Ginoux \cite{Gith}.} :

\begin{quote}

\og \textit{Condition de stabilit\'{e}}. -- Consid\'{e}rons donc une autre courbe non ferm\'{e}e satisfaisant
\`{a} l'\'{e}quation diff\'{e}rentielle, ce sera une sorte de spirale se
rapprochant ind\'{e}finiment de la courbe ferm\'{e}e. Si la courbe
ferm\'{e}e repr\'{e}sente un r\'{e}gime stable, en d\'{e}crivant la spirale
dans le sens de la fl\`{e}che on doit \^{e}tre ramen\'{e} sur la courbe
ferm\'{e}e, et c'est \`{a} cette seule condition que la courbe ferm\'{e}e
repr\'{e}sentera un r\'{e}gime stable d'ondes entretenues et donnera lieu
\`{a} la solution du probl\`{e}me. \fg{} \cite[p. 391]{P6}

\end{quote}

L'\'{e}quation diff\'{e}rentielle (J$_{1}$) \'{e}tablie par Janet \cite{Janet1919} est \'{e}galement incompl\`{e}te comme il le souligne d'ailleurs~:

\begin{quote}

\og Mais le ph\'{e}nom\`{e}ne est limit\'{e} par la courbure de la
caract\'{e}ristique, et en fait, il s'\'{e}tablit des oscillations
r\'{e}guli\`{e}res, non-sinuso\"{\i}dales, r\'{e}gies par l'\'{e}quation
(J$_{1}$) qu'on ne pourrait int\'{e}grer que si l'on connaissait la forme
explicite de la fonction $f\left( i \right)$. \fg{} \cite[p. 765]{Janet1919}

\end{quote}

En comparant cette derni\`{e}re citation de Janet avec la premi\`{e}re de Poincar\'{e} on constate que tous deux sont conscients du probl\`{e}me que constitue la repr\'{e}sentation math\'{e}matique de la \textit{caract\'{e}ristique d'oscillation} de ces dispositifs (machine s\'{e}rie-dynamo, arc chantant ou lampe \`{a} trois \'{e}lectrodes) qui fait appel \`{a} l'interpolation polyn\^{o}miale d'une courbe et qui rel\`{e}ve de ce qu'on appelle aujourd'hui la \textit{mod\'{e}lisation d'observables}. Ceci implique l'obtention d'un nombre minimum de points, $i.e.$, d'une s\'{e}rie de donn\'{e}es ou de mesures ce qui n\'{e}cessite d'une part la reproductibilit\'{e} de l'exp\'{e}rience \`{a} l'identique et, d'autre part d'avoir un appareil de mesure susceptible de fournir des valeurs suffisamment pr\'{e}cises. Cependant, en \'{e}tablissant dans cette note une analogie entre trois dispositifs diff\'{e}rents, Janet a montr\'{e} qu'ils relevaient du m\^{e}me ph\'{e}nom\`{e}ne oscillatoire dont il a fourni d'une part l'\'{e}quation, certes incompl\`{e}te, mais g\'{e}n\'{e}rale et, a d'autre part soulign\'{e} le fait que ces oscillations \'{e}taient \og non-sinusoïdales \fg.
De plus, il a mis en lumi\`{e}re la principale difficult\'{e} \`{a} surmonter pour compl\'{e}ter son \'{e}quation. En novembre de cette m\^{e}me ann\'{e}e 1919, c'est Blondel \cite{Blondel1919c} qui va r\'{e}soudre ce probl\`{e}me en \'{e}tablissant un an avant Van der Pol \cite{VdP1920} l'\'{e}quation de la triode.

\newpage

\section{Sur les oscillations auto-entretenues} \hfill

Vers 1919, peu apr\`{e}s la mort de son p\`{e}re, Blondel se remit curieusement \`{a} marcher \cite[p. 755, p. 26]{Bethenod1938,Broglie}. Durant cette p\'{e}riode Blondel \cite{Blondel1919a,Blondel1919b,Blondel1919c,Blondel1919d,Blondel1920,Blondel1923a,Blondel1923b} r\'{e}alisa toute une s\'{e}rie d'articles sur les oscillations entretenues par l'\textit{audion g\'{e}n\'{e}rateur}, c'est-\`{a}-dire, la \textit{lampe \`{a} trois \'{e}lectrodes}. Dans le premier d'entre eux intitul\'{e} \og Sur les syst\`{e}mes \`{a} oscillations persistantes, et en particulier sur les oscillations entretenues par auto-amorçage \fg, publi\'{e} en avril 1919 dans le \textit{Journal de Physique Th\'{e}orique et Appliqu\'{e}e}, Blondel \cite[p. 118]{Blondel1919a} introduisait la terminologie \og oscillations auto-entretenue \fg{}.

\subsection{Introduction de la terminologie} \hfill \\

Il est assez l\'{e}gitime de s'interroger sur l'origine de la terminologie \og auto-entretenue \fg{} et sur la date de son apparition dans la litt\'{e}rature scientifique. Des recherches entreprises dans les \textit{Comptes Rendus hebdomadaires des s\'{e}ances de l'Acad\'{e}mie des Sciences}, dans la \textit{Revue G\'{e}n\'{e}rale des Sciences Pures et Appliqu\'{e}es}, dans la \textit{Lumi\`{e}re \'{E}lectrique}, l'\textit{\'{E}clairage \'{E}lectrique} ainsi que dans le \textit{Journal de Physique Th\'{e}orique et Appliqu\'{e}e} n'ont pas permis, pour l'instant, de trouver une trace de ce n\'{e}ologisme avant 1919. La premi\`{e}re occurence semble \^{e}tre celle introduite par Blondel \cite{Blondel1919a}.

\subsection{Sur les oscillations entretenues par auto-amorçage} \hfill \\

Dans la premi\`{e}re partie de son m\'{e}moire, Blondel \cite{Blondel1919a} propose une nouvelle classification des oscillations selon trois types~:

\begin{enumerate}
\item[1.] {\sc oscillations entretenues proprement dites}
	\begin{enumerate}
	\item[a)] \textit{Oscillations entretenues par action discontinue.}
	\item[b)] \textit{Oscillations entretenues par action continue ou oscillations auto-entretenues.}
	\end{enumerate}
\item[2.] {\sc oscillations persistantes par \'{e}coulement fractionn\'{e}}
\item[3.] {\sc oscillations de longue p\'{e}riode produite par inversion de la force agissante}
\end{enumerate}

Blondel \cite[p. 117]{Blondel1919a} explique que le premier type (1.a) correspond aux oscillations entretenues telles qu'on les entend aujourd'hui et illustre son propos avec le pendule entretenu. En ce qui concerne le type (1.b) il choisit curieusement l'exemple des oscillations entretenues par un \textit{ronfleur}\footnote{Relais \'{e}lectromagn\'{e}tique \`{a} lame vibrante, dont le fonctionnement se traduit par un ronflement de basse fr\'{e}quence, qui remplace parfois une sonnette dans les installations t\'{e}l\'{e}phoniques.}. Il pr\'{e}cise alors que le condensateur de capacit\'{e} $C$  que comporte ce circuit peut \^{e}tre repr\'{e}sent\'{e} par une antenne de T.S.F. jouant un r\^{o}le analogue. Peut-\^{e}tre s'agit-il l\`{a} d'une mani\`{e}re de se rattacher au domaine de la Radiotechnique. Blondel ajoute ensuite :

\begin{quote}

\og \ldots on constate ais\'{e}ment que le circuit oscillant pourra entretenir lui-m\^{e}me ses oscillations, si le courant qui le parcourt provoque d'une mani\`{e}re convenable la production d'un courant alternatif ou d'une force \'{e}lectromotrice alternative dans une source \'{e}trang\`{e}re disposant par elle-m\^{e}me d'une \'{e}nergie suffisante pour que le circuit oscillant reçoive plus d'\'{e}nergie du courant ainsi provoqu\'{e} qu'il n'en d\'{e}pensera lui-m\^{e}me dans le m\'{e}canisme provoquant la formation de ces courant auxiliaires. \fg{}\\ \vphantom{ } \hfill \cite[p. 120]{Blondel1919a}

\end{quote}

Il est int\'{e}ressant de comparer cette phrase avec celle que prononcera Balthazar Van der Pol lors de sa seconde conf\'{e}rence le 11 mars 1930 \`{a} l'\'{E}cole Sup\'{e}rieure d'\'{E}lectricit\'{e} \`{a} Paris :

\begin{quote}

\og Une oscillation de relaxation a lieu, d'une mani\`{e}re g\'{e}n\'{e}rale, toutes les fois qu'un m\'{e}canisme, contenant une source d'\'{e}nergie continue, permet \`{a} un ph\'{e}nom\`{e}ne essentiellement ap\'{e}riodique de se r\'{e}p\'{e}ter automatiquement un nombre ind\'{e}fini de fois. \fg{} \cite[p. 300]{VdP1930}

\end{quote}

Blondel met alors en \'{e}vidence la cause responsable de l'entretien de ces oscillations :

\begin{quote}

\og \ldots si, au contraire, l'apport d'\'{e}nergie est plus fort que la d\'{e}pense, \textit{l'amortissement deviendra n\'{e}gatif} : il y aura entretien des oscillations et tendances de celles-ci \`{a} s'accro\^{i}tre avec le temps, d'autant plus que la valeur absolue de l'amortissement n\'{e}gatif sera grande. Si cette condition est remplie pour les oscillations voisines de z\'{e}ro, elles seront \textit{auto-amorçantes}. \fg{} \cite[p. 120]{Blondel1919a}

\end{quote}

Il fait ici implicitement r\'{e}f\'{e}rence au concept de \textit{r\'{e}sistance n\'{e}gative} introduit par Hans Luggin \cite[p. 568]{Luggin} pour expliquer l'entretien des oscillations d'un arc chantant. On peut \'{e}galement mettre cette phrase en perspective avec celle \'{e}nonc\'{e}e par Van der Pol lors de son premier expos\'{e} \`{a} Paris le 24 mai 1928 dans la Salle de la Soci\'{e}t\'{e} de G\'{e}ographie :

\begin{quote}

\og Mais en \'{e}lectrotechnique et particuli\`{e}rement dans le domaine de radio-t\'{e}l\'{e}graphie nous avons \`{a} notre disposition des r\'{e}sistances d'un caract\`{e}re n\'{e}gatif. Au lieu de dissiper de l'\'{e}nergie ces r\'{e}sistances peuvent \textit{fournir} de l'\'{e}nergie. Il est \'{e}vident par cons\'{e}quent qu'elles ne se pr\'{e}sentent que dans les syst\`{e}mes qui contiennent une source d'\'{e}nergie. \fg{} \cite[p. 367]{VdP1928}

\end{quote}

Dans le second type (2.), Blondel cite les exemples de l'arc musical de Duddell, de l'arc de \textit{seconde esp\`{e}ce} qu'il a \og d\'{e}couvert \fg{} en 1905 (voir Blondel \cite{Blondel1905a,Blondel1905b,Blondel1905c}) et du vase de Tantale ou du vase culbuteur\footnote{Il s'agit d'un r\'{e}cipient mont\'{e} sur un axe pivotant plac\'{e} en son centre et qui bascule brutalement lorsque le niveau de remplissage a rompu l'\'{e}quilibre. Voir Van der Pol \cite[p. 307]{VdP1930} et Le Corbeiller \cite[p. 6 et p. 42]{LeCorbeiller1931}.}.\\

Blondel illustre le troisi\`{e}me type (3.) d'oscillations \`{a} partir de l'exp\'{e}rience de la machine s\'{e}rie-dynamo d\'{e}crite par Janet \cite{Janet1919} dans sa note \og Sur une analogie \'{e}lectrotechnique \fg{} dont il rappelle le principe et, aussi \`{a} partir des travaux d'Henry L\'{e}aut\'{e} \cite{Léauté} concernant les machines hydrauliques munies d'un r\'{e}gulateur \`{a} action indirecte. Ce dernier avait en effet employ\'{e}, pour \'{e}tudier la stabilit\'{e} des oscillations \og \`{a} longues p\'{e}riodes \fg{} au sein de ces dispositifs, un diagramme comportant position en abscisse et vitesse en ordonn\'{e}e correspondant exactement \`{a} une repr\'{e}sentation dans le \textit{plan de phase} de Poincar\'{e}. Il avait alors obtenu des courbes ferm\'{e}es en tous points similaires \`{a} des \textit{cycles limites} de Poincar\'{e}. Dans l'\'{e}dition russe ainsi que dans l'\'{e}dition anglo-saxonne de l'ouvrage d'Andronov \textit{et al.} \cite{Andronov1959} remani\'{e} par son \'{e}l\`{e}ve N.A. Zheleztsov, il rappelait dans une note de bas de page du second paragraphe du premier chapitre \cite[p. 39]{Andronov1959} que L\'{e}aut\'{e} avait \'{e}t\'{e} le premier \`{a} faire usage du concept de \textit{plan de phase} pour d\'{e}crire les courbes int\'{e}grales ferm\'{e}es de ces machines hydrauliques sans cependant r\'{e}aliser qu'il s'agissait de \textit{cycles limites} de Poincar\'{e}.\\

Dans la derni\`{e}re partie de son expos\'{e} Blondel \cite[p. 153]{Blondel1919a} va mettre en \'{e}vidence la distinction entre \textit{oscillations entretenues} et \textit{oscillations auto-entretenues} \`{a} partir de nombreux exemples illustr\'{e}s avec un \'{e}tonnant diagramme qui consiste \`{a} repr\'{e}senter l'amplitude des oscillations sous la forme trigonom\'{e}trique d'un nombre complexe :

\begin{quote}

\og La diff\'{e}rence entre les oscillations entretenues par action discontinue et celles entretenues par action continue peut \^{e}tre pr\'{e}cis\'{e}e par un diagramme en remarquant que cette oscillation amortie peut s'\'{e}crire sous la forme~:

\smallskip

\hfill $\rho =Ae^{\left( {-\alpha +j\beta } \right)t}$, \hfill \null

\smallskip

qui se pr\^{e}te \`{a} une repr\'{e}sentation graphique comme toutes les fonctions imaginaires (\textit{fig. 8}).

\begin{figure}[htbp]
\centerline{\includegraphics[width=8cm,height=8cm]{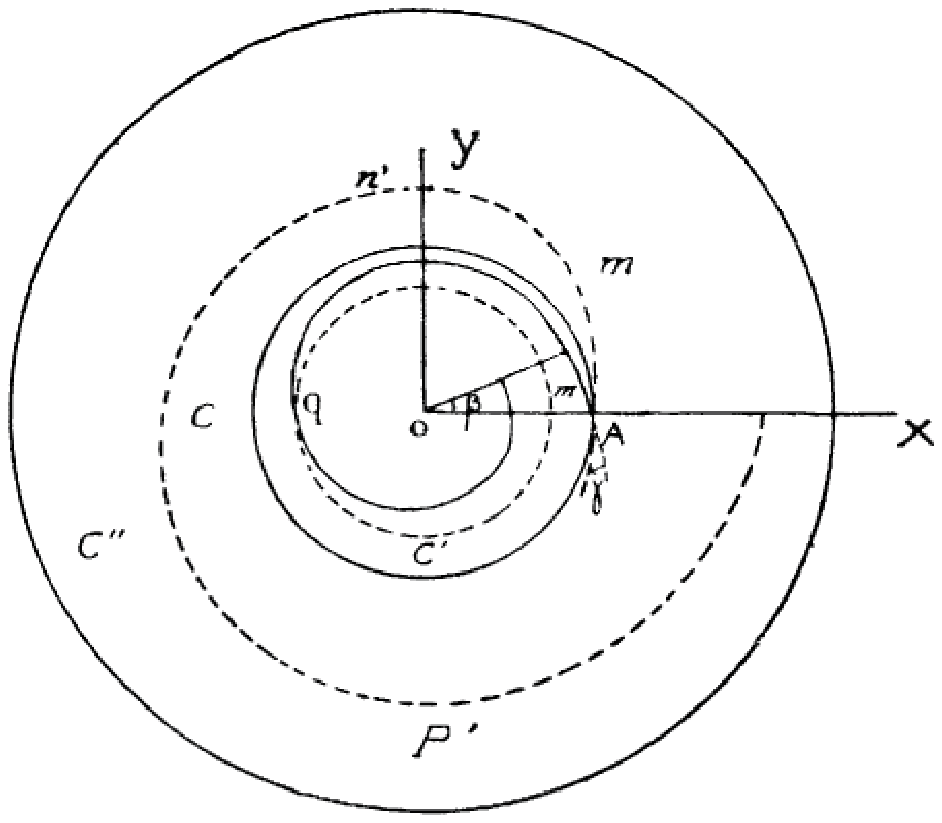}}
\caption{ }
\label{fig8}
\end{figure}

Si l'on trace $\rho$ en coordonn\'{e}es polaires et d\'{e}signe par $\beta t$ l'angle d\'{e}crit \`{a} partir d'un axe OX par le rayon figuratif $\rho $, le point mobile $m$, extr\'{e}mit\'{e} de ce rayon, d\'{e}crit, comme on le sait, une spirale logarithmique qui se r\'{e}duirait au cercle de rayon OA si l'amortissement \'{e}tait nul~;~l'amortissement donne~\`{a} la tangente de la courbe, par rapport au rayon un coefficient angulaire $-\dfrac{\beta }{\alpha }$. \fg{} \\ \vphantom{} \hfill \cite[p. 153]{Blondel1919a}

\end{quote}

\newpage

Il consid\`{e}re alors tout d'abord le cas des oscillations entretenues par action discontinue, $i.e.$ des \textit{oscillations forc\'{e}es} pour lesquelles il \'{e}tudie graphiquement \og l'effet de l'impulsion d'entretien \fg{} sur la forme du diagramme (voir Fig. 1) avant de s'int\'{e}resser \`{a} celui des oscillations entretenues
par action continue, $i.e.$ des \textit{oscillations auto-entretenues}.

\begin{quote}

\og Th\'{e}oriquement l'entretien continu doit fournir au mobile (ou \`{a} la
variable \'{e}lectrique qui en tient lieu dans les oscillations
\'{e}lectriques) une \'{e}nergie compensant l'\'{e}nergie d\'{e}grad\'{e}e
par l'amortissement spontan\'{e} du syst\`{e}me abandonn\'{e} \`{a}
lui-m\^{e}me. L'entretien id\'{e}al serait donc celui qui aurait pour effet
de substituer \`{a} la spirale le cercle C, qui a m\^{e}me rayon initial.
Mais un tel syst\`{e}me ne pr\'{e}senterait pas de stabilit\'{e}, tout au
moins dans l'hypoth\`{e}se o\`{u} l'amortissement $\alpha $ est
ind\'{e}pendant de l'amplitude. \fg{} \\ \vphantom{} \hfill \cite[p. 154]{Blondel1919a}

\end{quote}

Il souligne ici \`{a} nouveau la n\'{e}cessit\'{e} de compenser l'\'{e}nergie dissip\'{e}e par l'amortissement pour que le syst\`{e}me s'auto-entretienne et indique les transformations que cela produirait au niveau du diagramme. Bien qu'il ne s'agisse pas d'un portrait de phase le passage d'une spirale \`{a} un cercle, $i.e.$ la modification du comportement de
la solution en fonction de la valeur d'un param\`{e}tre (l'amortissement $\alpha $ dans ce cas) et par cons\'{e}quent le changement de stabilit\'{e} correspond \`{a} ce que Poincar\'{e} \cite[p. 261]{P8} avait appel\'{e} une \textit{bifurcation}. Blondel pr\'{e}cise alors~:

\begin{quote}

\og Pour qu'il y ait stabilit\'{e}, il est n\'{e}cessaire que le ph\'{e}nom\`{e}ne d'entretien donne au syst\`{e}me oscillant un
amortissement tr\`{e}s petit, mais l\'{e}g\`{e}rement positif, c'est-a-dire que $-\alpha $ soit remplac\'{e} par $+\alpha $ et la spirale $Amq$ de rayon d\'{e}croissant est remplac\'{e}e par une autre spirale logarithmique $A{m}'{n}'{p}'$. Les oscillations seront alors du type divergent et l'amplitude cro\^{\i}trait ind\'{e}finiment si rien n'intervenait pour la limiter. Mais, en g\'{e}n\'{e}ral, il y a toujours dans des ph\'{e}nom\`{e}nes de ce genre une action limitatrice consistant, soit dans
une diminution de l'\'{e}nergie motrice, soit dans un accroissement de la r\'{e}sistance passive, c'est-\`{a}-dire de l'\'{e}nergie d\'{e}grad\'{e}e quand augmente l'amplitude. La spirale logarithmique se trouve ainsi limit\'{e}e d'elle-m\^{e}me \`{a} un troisi\`{e}me cercle $C''$ dont elle ne peut sortir. \fg{} \cite[p. 154]{Blondel1919a}

\end{quote}

\newpage

S'il semble que la description de Blondel pour ces \og cercles dont on ne peut sortir \fg{} ressemble beaucoup \`{a} celle de Poincar\'{e} on ne peut cependant pas les qualifier de \og cycles limites \fg{}. Pour s'en convaincre il suffit de se reporter \`{a} la \textit{Notice sur les Travaux scientifiques d'Henri Poincar\'{e}} faite par lui-m\^{e}me en 1886 et dans laquelle il d\'{e}finissait ainsi ce concept~:

\begin{quote}

\og [{\ldots}] il y a un autre genre de courbes ferm\'{e}es qui jouent un r\^{o}le capital dans cette th\'{e}orie~: ce sont les \textit{cycles limites}. J'appelle ainsi les courbes ferm\'{e}es qui satisfont \`{a} notre \'{e}quation diff\'{e}rentielle et dont les autres courbes d\'{e}finies par la m\^{e}me \'{e}quation se rapprochent asymptotiquement sans jamais les atteindre.
Cette seconde notion n'est pas moins importante que la premi\`{e}re. Supposons, en effet, que l'on ait trac\'{e} un cycle limite~; il est clair que le point mobile dont nous parlions plus haut ne pourra jamais le franchir et qu'il restera toujours \`{a} l'int\'{e}rieur de ce cycle, ou toujours \`{a} l'ext\'{e}rieur. \fg{} \cite[p. 30]{P9}

\end{quote}

Il appara\^{i}t donc que deux conditions sont n\'{e}cessaires pour recouvrir ce concept. La premi\`{e}re est de tendre asymptotiquement vers ces cycles limites et la seconde de ne jamais les franchir. Malheureusement, la description de Blondel ne satisfait qu'\`{a} la derni\`{e}re de ces deux conditions et il semble avoir laiss\'{e} de c\^{o}t\'{e} l'aspect attractif
de ces cycles.

Blondel consid\`{e}re ensuite le cas des oscillations entretenues par une lampe \`{a} trois \'{e}lectrodes qu'il d\'{e}crit au moyen du m\^{e}me type de repr\'{e}sentation graphique (voir Fig. 9). Il explique alors \`{a} partir de la figure le ph\'{e}nom\`{e}ne~:

\begin{figure}[htbp]
\centerline{\includegraphics[width=9.1cm,height=7.83cm]{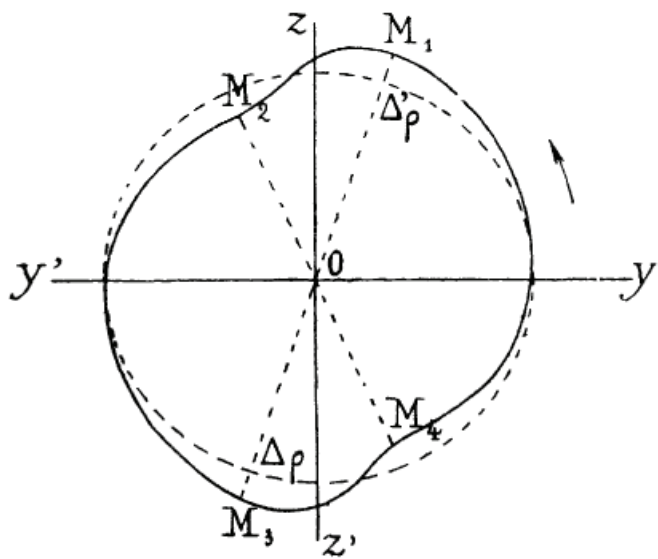}}
\caption[Oscillations entretenues par une lampe \`{a} trois \'{e}lectrodes]{Oscillations entretenues par une lampe \`{a} trois\\ \centering{\'{e}lectrodes, d'apr\`{e}s Blondel \cite[p. 157]{Blondel1919a}}}
\label{fig9}
\end{figure}

\begin{quote}

\og Le syst\`{e}me est auto-amorçant. L'extr\'{e}mit\'{e} du rayon
vecteur $\rho $ d\'{e}crit ainsi une spirale logarithmique centrifuge.
Arriv\'{e} en M$_{1}$ au voisinage du maximum de l'amplitude de
l'oscillation, le r\'{e}gime est troubl\'{e} pendant un instant dans l'angle
M$_{1}$OM$_{2}$, il devient amorti ($\alpha <0$) \`{a} cause de la courbure
des caract\'{e}ristiques de la lampe qui correspond \`{a} un accroissement
de la r\'{e}sistance int\'{e}rieure de celle-ci ; le vecteur $\rho $ est
diminu\'{e} d'apr\`{e}s une loi compliqu\'{e}e. Supposons qu'en M$_{2}$ le
r\'{e}gime soit de nouveau entretenu ; \`{a} partir de ce point, $\rho $
d\'{e}crit une nouvelle portion de spirale logarithmique centrifuge ($\alpha
>0$). En r\'{e}sum\'{e}, l'amplitude tend constamment \`{a} cro\^{i}tre,
mais subit des freinages p\'{e}riodiques \`{a} la fin de chaque alternance.
La condition de stabilit\'{e} autour du cercle C exige que les impulsions
$\Delta \rho $ p\'{e}riodiques soient n\'{e}gatives et que leur valeur
absolue croisse quand $\rho $ augmente. C'est ainsi que fonctionnent les
audions g\'{e}n\'{e}rateurs quand le diagramme est appliqu\'{e} \`{a}
repr\'{e}senter leurs vecteurs d'intensit\'{e} du courant de plaque, car la
r\'{e}sistance int\'{e}rieure qui amortit le courant s'accro\^{i}t quand on
s'\'{e}carte des parties rectilignes des caract\'{e}ristiques ; le
r\'{e}gime oscillatoire amorti peut \^{e}tre consid\'{e}r\'{e} comme
recevant pendant un court instant une impulsion n\'{e}gative n\'{e}cessaire
pour amener l'oscillation du courant \`{a} sa valeur moyenne. \fg{} \cite[p. 157]{Blondel1919a}

\end{quote}

Tout d'abord on peut remarquer qu'il est le premier \`{a} qualifier ce
syst\`{e}me \og d'auto-amorçant \fg{} ce qui peut s'entendre dans son
acception par \og auto-entretenu \fg{}. Cependant, m\^{e}me si le sch\'{e}ma
ainsi que les explications qu'il fournit donnent \`{a} penser qu'il est en
pr\'{e}sence d'un cycle limite de Poincar\'{e} il ne fait pas le lien avec
ses travaux.\\

Enfin, Blondel va d\'{e}crire toujours avec le m\^{e}me type de
repr\'{e}sentation graphique les oscillations entretenues produites par
l'arc musical de Duddell et l'arc strident intermittent qu'il avait
d\'{e}couvert quinze ans auparavant.

\begin{figure}[htbp]
\centerline{\includegraphics[width=8.0465cm,height=7.1155cm]{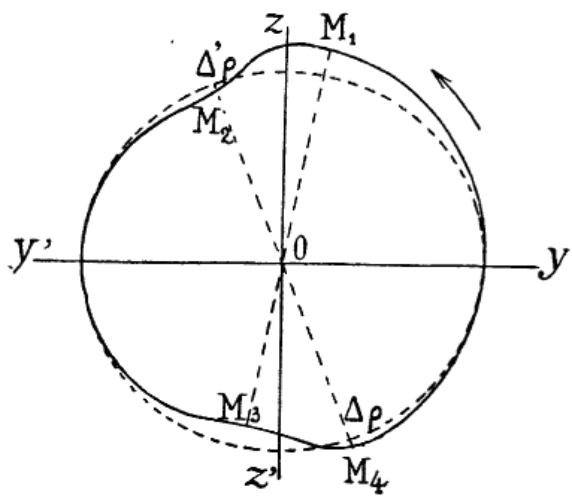}}
\caption[Oscillations entretenues par l'arc musical de Duddell]{Oscillations entretenues par l'arc musical de\\ \centering{Duddell, d'apr\`{e}s Blondel \cite[p. 160]{Blondel1919a}}}
\label{fig10}
\end{figure}

\newpage

Blondel d\'{e}crit en ces termes le ph\'{e}nom\`{e}ne:

\begin{quote}

\og Les conditions de stabilit\'{e} d'amplitude seront \'{e}videmment les
suivantes : tant que l'amplitude $\rho $ sera inf\'{e}rieure \`{a} sa valeur
normale, l'impulsion positive, pendant le r\'{e}gime troubl\'{e}
acc\'{e}l\'{e}rateur M$_{1}$M$_{2}$, devra \^{e}tre plus grande en valeur
absolue que l'impulsion n\'{e}gative entre M$_{3}$ et M$_{4}$, $\Delta \rho
>\left( {-{\Delta }'\rho } \right)$. Au contraire, lorsque l'amplitude, pour
une cause quelconque, aura d\'{e}pass\'{e} le cercle C, il faudra qu'elle
tende \`{a} y revenir, ce qui se traduit par la condition $\left( {-{\Delta}'\rho } \right)>\Delta \rho $. \fg{} \cite[p. 160]{Blondel1919a}

\end{quote}

Dans le cas de l'arc musical de Duddell il pr\'{e}cise que l'amplitude devra tendre \`{a} revenir vers le cercle C qui constitue ainsi un \og attracteur \fg{} pour l'amplitude et qui satisfait \textit{de facto} une partie de la premi\`{e}re condition de Poincar\'{e}. En effet, en \'{e}tudiant la stabilit\'{e} de l'amplitude des oscillations Blondel veut montrer qu'elle est born\'{e}e ce qui se traduit g\'{e}om\'{e}triquement par cette \og tendance \`{a} revenir vers le cercle C \fg{}. Si Blondel a pu mettre en \'{e}vidence le caract\`{e}re attractif de ces \og cercles dont on ne peut sortir \fg{} ou vers lesquels on a \og tendance \`{a} revenir \fg{}, recouvrant ainsi la seconde condition introduite par Poincar\'{e} pour d\'{e}finir le concept de \og cycles limites \fg{}, la premi\`{e}re qui implique de tendre asymptotiquement vers ces cycles lui est inaccessible. Non pas pour des raisons th\'{e}oriques mais plut\^{o}t \`{a} cause du point de vue qu'il a choisi pour repr\'{e}senter le ph\'{e}nom\`{e}ne, $i.e.$ celui d'une mise sous \og forme normale complexe \fg{}. En effet, Blondel cherche \`{a} montrer g\'{e}om\'{e}triquement la stabilit\'{e} de l'amplitude c'est pourquoi il choisit d\'{e}lib\'{e}r\'{e}ment une condition initiale tr\`{e}s proche du cercle dans le cas de la lampe \`{a} trois \'{e}lectrodes :

\begin{quote}

\og Partons du point M qui correspond \`{a} une amplitude nulle de
l'oscillation du courant du circuit de plaque. \fg{} \\ \vphantom{} \hfill \cite[p. 157]{Blondel1919a}

\end{quote}

comme dans celui de l'arc musical de Duddell~:

\begin{quote}

\og Partant d'un point d'\'{e}longation nulle (courant nul). \fg{} \\ \vphantom{} \hfill \cite[p. 169]{Blondel1919a}

\end{quote}

Ainsi, s'il a pu montrer qu'une condition initiale d'amplitude nulle ne
s'\'{e}loignait pas du cercle C, il n'a pas envisag\'{e} le cas de
conditions initiales d'amplitude non-nulle qui tendraient asymptotiquement
\`{a} y revenir. Ses repr\'{e}sentations constituent cependant bien des
\og cycles limites \fg{} de Poincar\'{e} m\^{e}me s'il ne les a pas
identifi\'{e}es comme tels.

\section{Mod\'{e}lisation de la caract\'{e}ristique d'oscillation}

Comme l'avait soulign\'{e} Poincar\'{e} \cite{P6}, puis Janet \cite{Janet1919}, le principal obstacle \`{a} la mise en \'{e}quation compl\`{e}te des oscillations observ\'{e}es dans la machine s\'{e}rie-dynamo, l'arc chantant et la lampe \`{a} trois \'{e}lectrodes \'{e}tait la mod\'{e}lisation de la caract\'{e}ristique d'oscillation du composant non-lin\'{e}aire, analogue \`{a} une r\'{e}sistance n\'{e}gative, pr\'{e}sent dans ces trois dispositifs.

\subsection{Oscillations entretenues par une lampe \`{a} trois \'{e}lectrodes} \hfill

Dans une note publi\'{e}e aux $C.R.A.S.$ et pr\'{e}sent\'{e}e \`{a} l'Acad\'{e}mie des Sciences le 17 novembre 1919, Blondel surmonte cette difficult\'{e} en mod\'{e}lisant la caract\'{e}ristique d'oscillations de la lampe \`{a} trois \'{e}lectrodes. En appelant $u$ la tension de la plaque \`{a} la variation $i$ du courant de plaque de la lampe \`{a} trois \'{e}lectrodes, $v$ le potentiel de grille et $k$ le coefficient d'amplification de la lampe \`{a} trois \'{e}lectrodes il explicite la relation entre $u$ et $i~$ dans la lampe \`{a} trois \'{e}lectrodes qui \`{a} la forme $i=F\left( {u+kv} \right)$ et \og dans laquelle $F$ repr\'{e}sente une fonction qui se traduit par une courbe connue pr\'{e}sentant une tr\`{e}s longue inflexion aux environs de la valeur moyenne du courant statique I (\'{e}gale \`{a} environ la moiti\'{e} du courant de saturation). \fg{} \cite[p. 946]{Blondel1919c}. Il fait alors l'hypoth\`{e}se que l'on ne sort pas de la r\'{e}gion o\`{u} cette courbe conserve la m\^{e}me forme quel que soit $v$ et se d\'{e}place seulement parall\`{e}lement \`{a} elle-m\^{e}me par une translation suivant l'axe des $u$ quand on fait varier $v$. Ceci le conduit \`{a} mod\'{e}liser la caract\'{e}ristique d'oscillation $i=F\left( {u+kv} \right)$ de la lampe \`{a} trois \'{e}lectrodes en proposant \og de la d\'{e}velopper sous forme d'une s\'{e}rie \`{a} termes impairs, qui sera s\^{u}rement convergente \fg{} \cite[p. 946]{Blondel1919c}. Il obtient ainsi~:

\smallskip

\begin{center}

\hfill
$i=F\left( {u+kv} \right)=b_1 \left( {u+kv} \right)-b_3 \left(
{u+kv} \right)^3-b_5 \left( {u+kv} \right)^5-\ldots $ \hfill (B$_{1}$)

\end{center}

\smallskip

Le but de cette note intitul\'{e}e \og Amplitude du courant oscillant produit par les audions g\'{e}n\'{e}rateurs \fg{} est en fait de calculer une approximation de l'amplitude des oscillations. C'est pour cette raison que Blondel cherche \`{a} \'{e}tablir l'\'{e}quation diff\'{e}rentielle des oscillations de la lampe \`{a} trois \'{e}lectrodes. Aussi, en appelant $i$ l'intensit\'{e} du courant de plaque \`{a} l'instant $t$, $i_1$ et $i_2$ les intensit\'{e}s dans les branches de self-induction $L$ et de capacit\'{e} $C$, ayant pour r\'{e}sistances internes $r_1$ et $r_2$ respectivement, $u$ la tension oscillante aux bornes des circuits d\'{e}riv\'{e}s, il obtient les trois \'{e}quations suivantes

\[
i_1 +i_2 =i\mbox{,} \quad r_1 i_1 +L\frac{di_1 }{dt}=u\mbox{,} \quad r_2 i_2 +\frac{1}{C}\int{i_2 dt} =u\mbox{,} \quad h=\frac{kM}{L}-1;
\]

o\`{u} $M$ repr\'{e}sente le coefficient d'induction mutuelle et o\`{u} le potentiel de grille $v \approx - \dfrac{M}{L}u$.
En les combinant et en les d\'{e}rivant Blondel \cite[p. 945]{Blondel1919c} parvient \`{a} \'{e}tablir l'\'{e}quation diff\'{e}rentielle de la lampe \`{a} trois \'{e}lectrodes sous la forme :\\

\begin{center}

\hfill $\dfrac{d^3u}{dt^3}+\dfrac{r_2 }{L}\dfrac{d^2u}{dt^2}+\left(
{\dfrac{1}{CL}-\dfrac{r_1 r_2 }{L^2}} \right)\dfrac{du}{dt}-\dfrac{r_1
}{CL^2}u-r_2 \dfrac{d^3i}{dt^3}-\dfrac{1}{C}\dfrac{d^2i}{dt^2}=0$ \hfill (6)

\end{center}

\smallskip

La pr\'{e}sence des r\'{e}sistances internes $r_1$ et $r_2$ que Blondel aurait pu n\'{e}gliger le conduit \`{a} cette \'{e}quation
diff\'{e}rentielle du troisi\`{e}me ordre et du second degr\'{e}. En substituant l'expression (B$_{1}$) de l'intensit\'{e} $i$ dans cette \'{e}quation (6) \og l'\'{e}quation finale du probl\`{e}me \fg{} \cite[p. 947]{Blondel1919c} prend la forme suivante~:\\

\begin{center}

\hfill $\begin{array}{l}
 \dfrac{d^3u}{dt^3}+\dfrac{d^2u}{dt^2}\dfrac{r_2 }{L}+\dfrac{du}{dt}\left(
{\dfrac{1}{CL}-\dfrac{r_1 r_2 }{L^2}} \right)-\dfrac{r_1 }{CL^2}u \\
 \mbox{ }-\left( {r_2 \dfrac{d^2u}{dt^2}-\dfrac{1}{C}\dfrac{d^2u}{dt^2}}
\right)\left[ {b_1 \left( {h-\dfrac{kMr_1 }{L^2}u} \right)-3b_3 h^3u^2-\ldots
} \right] \\
 \mbox{ }-\left[ {3r_2 \dfrac{d^3u}{dt^3}\dfrac{du}{dt}+\dfrac{1}{C}\left(
{\dfrac{du}{dt}} \right)^2} \right]\mbox{ }\left[ {-\dfrac{b_1 kMr_1
}{L^2}-6b_3 h^3u+\ldots } \right] \\
 \mbox{ }-r_2 \left( {\dfrac{du}{dt}} \right)^3\left[ {-6b_3 h^3-\ldots
\ldots \ldots \ldots \ldots \ldots \ldots \ldots \ldots \ldots \ldots} \right]=0 \\
 \end{array}$ \hfill (12)

\end{center}

En n\'{e}gligeant les r\'{e}sistances internes $r_1 $ et $r_2 $, $i.e.$ en posant dans l'Eq. (12)~: $r_1 =r_2 =0$, on a :

\[
\frac{d^3u}{dt^3}+\frac{du}{dt}\left( {\frac{1}{CL}}
\right)+\frac{1}{C}\frac{d^2u}{dt^2}\left[ {b_1 h-3b_3 h^3u^2-\ldots }
\right]-\frac{1}{C}\left( {\frac{du}{dt}} \right)^2\left[ {-6b_3 h^3u+\ldots
} \right]=0
\]

En regroupant les termes en $b_3 h^3$, on obtient~:

\[
\frac{d^3u}{dt^3}+\frac{du}{dt}\left( {\frac{1}{CL}}
\right)+\frac{1}{C}\frac{d^2u}{dt^2}\left( {b_1 h} \right)-\frac{3b_3
h^3}{C}\left[ {\frac{d^2u}{dt^2}u^2+\left( {\frac{du}{dt}} \right)^2\left(
{2u} \right)+\ldots } \right]=0
\]

En remarquant alors que le dernier terme s'\'{e}crit~:

\[
\left[ {\frac{d^2u}{dt^2}u^2+\left( {\frac{du}{dt}} \right)^2\left( {2u}
\right)+\ldots } \right]=\frac{d}{dt}\left( {\frac{du}{dt}u^2}
\right)+\ldots
\]

En int\'{e}grant une fois par rapport au temps, l'Eq. (12) devient~:

\smallskip

\begin{center}

\hfill $C\dfrac{d^2u}{dt^2}-\left( {b_1 h-3b_3 h^3u^2-\ldots } \right)\dfrac{du}{dt}+\dfrac{u}{L}=0$ \hfill (B$_{2}$)

\end{center}

En d\'{e}veloppant dans l'\'{e}quation (12) la tension $u$ en s\'{e}rie de Fourier et en identifiant terme \`{a} terme, Blondel d\'{e}duit une premi\`{e}re approximation de la p\'{e}riode plus exactement de la fr\'{e}quence angulaire ou pulsation $\omega $~des oscillations~:

\begin{center}

\hfill $\omega ^2\approx \dfrac{1}{CL}\left( {1+\dfrac{r_2 }{\rho }h_0 -r_1 r_2
\dfrac{C}{L}} \right)$ avec $h_0 =\rho C\dfrac{r_1 +r_2 }{L}$  \hfill (15)

\end{center}

Si l'on n\'{e}glige \`{a} nouveau les r\'{e}sistances internes $r_1 $ et $r_2 $ cette expression se r\'{e}duit \`{a}~:

\begin{center}

\hfill $\omega ^2\approx \dfrac{1}{CL}$ \hfill (B$_{3}$)

\end{center}

\vphantom{} \smallskip

Il est important de remarquer que Blondel retrouve la formule de Thomson en premi\`{e}re approximation. Avec la m\^{e}me
technique, il obtient une premi\`{e}re approximation de l'amplitude A$_{1}$ des oscillations~:\\

\begin{center}

\hfill $A_1 \approx \dfrac{2}{h}\sqrt {\dfrac{-b_1 +\dfrac{r_1 +r_2 }{hL}C}{-3b_3
}}$ \hfill (16)

\end{center}

\vphantom{} \smallskip

En n\'{e}gligeant encore les r\'{e}sistances internes $r_1$ et $r_2$ cette expression se r\'{e}duit \`{a}~:

\begin{center}

\hfill $A_1 \approx \dfrac{1}{h}\sqrt {\dfrac{4}{3} \dfrac{b_1 }{b_3 }} $ \hfill (B$_{4}$)

\end{center}

\vphantom{} \smallskip

Au mois de juin de l'ann\'{e}e suivante, Blondel \cite{Blondel1920} va exposer ces r\'{e}sultats dans un article plus long et plus d\'{e}taill\'{e}, publi\'{e} dans la revue \textit{Radio\'{e}lectricit\'{e}}.

\subsection{Oscillations entretenues par un arc chantant (I)} \hfill

\vspace{0.1in}

Entr\'{e} \`{a} dix-huit ans \`{a} l'\'{E}cole Polytechnique, Pomey (1861-1943) int\`{e}gre ensuite l'\'{E}cole Sup\'{e}rieure des Postes et T\'{e}l\'{e}graphes.\\ Il est promu Ing\'{e}nieur des T\'{e}l\'{e}graphes en 1883 et remplit cette fonction \`{a} Clermont-Ferrand, au Mans, \`{a} Nice, \`{a} Tours et \`{a} Ch\^{a}lons-sur-Marne. Il passe ensuite une licence en droit puis une licence \`{e}s sciences. En 1893, il est nomm\'{e} professeur d'\'{e}lectricit\'{e} th\'{e}orique \`{a} l'\'{E}cole Sup\'{e}rieure des Postes et T\'{e}l\'{e}graphes \og concurremment \fg{} avec Poincar\'{e}.\\ Il y enseignera pendant trente ann\'{e}es durant lesquelles ses cours seront r\'{e}guli\`{e}rement publi\'{e}s (Voir Pomey \cite{Pomey1914,Pomey1928,Pomey1931}) ainsi que les conf\'{e}rences qu'il r\'{e}alisa au sein de cette \'{E}cole dont il deviendra directeur de 1924 \`{a} 1926 avant de prendre sa retraite en 1927.

Durant la premi\`{e}re guerre mondiale, il est nomm\'{e} Chef du service T\'{e}l\'{e}graphique de la place de Toul puis de celui de l'Arm\'{e}e d'Orient. Il est alors l'Adjoint du Directeur du Service de la Radiot\'{e}l\'{e}graphie militaire : le Colonel Gustave Ferri\'{e}. Au sortir du conflit il reprend ses activit\'{e}s et ses publications. Le 28 juin 1920 para\^{i}t son ouvrage intitul\'{e} : \og Introduction \`{a} la th\'{e}orie des courants t\'{e}l\'{e}phoniques et de la radiot\'{e}l\'{e}graphie\footnote{La date exacte de la parution a \'{e}t\'{e} pr\'{e}cis\'{e}e par Madame Christine Robert (Administrateur du Pr\^{e}t Interbiblioth\`{e}ques de la Biblioth\`{e}que Nationale de France) en consultant l'ouvrage : la Bibliographie de la France, volume ann\'{e}e 1920, journal de la librairie.} \fg{} \cite{Pomey1920} pr\'{e}fac\'{e} par Andr\'{e} Blondel. Le chapitre XIX consacr\'{e} \`{a} la \og G\'{e}n\'{e}ration des oscillations entretenues \fg{} d\'{e}bute ainsi :

\begin{quote}

\og Pour que les oscillations entretenues subsistent dans un syst\`{e}me, il faut qu'elles correspondent \`{a} un r\'{e}gime stable ne d\'{e}pendant que de l'\'{e}tat du syst\`{e}me ou de ses variations p\'{e}riodiques, mais invariable quand on ne change que dans certaines limites les conditions initiales qui y ont donn\'{e} naissance. Il en r\'{e}sulte que le r\'{e}gime p\'{e}riodique permanent se pr\'{e}sente comme une solution asymptotique pour $t$ infini. \fg{} \cite[p. 373]{Pomey1920}

\end{quote}

Il est int\'{e}ressant de comparer cette phrase d'une part avec la \textit{condition de stabilit\'{e}} \'{e}nonc\'{e}e en mai-juin 1908 par Henri Poincar\'{e} \cite[p. 391]{P6} (voir p. 9) dans ses conf\'{e}rences donn\'{e}es \`{a} l'\'{E}cole Sup\'{e}rieure des Postes et T\'{e}l\'{e}graphes et, d'autre part avec cette phrase qu'Aleksandr Andronov (1901-1952) \'{e}crira moins de dix ans plus tard dans dans sa c\'{e}l\`{e}bre note aux \textit{Comptes Rendus}\footnote{Andronov \cite{Andronov1928} avait d\'{e}j\`{a} pr\'{e}sent\'{e} ce r\'{e}sultat lors du VI\ieme{} congr\`{e}s des physiciens russes qui eut lieu \`{a} Moscou entre le 5 et 16 ao\^{u}t 1928.} :

\begin{quote}

\og Il est donc clair que la p\'{e}riode et l'amplitude des oscillations stationnaires ne d\'{e}pendent pas des conditions initiales. \fg{} \cite[p. 560]{Andronov1929}

\end{quote}

Il appara\^{i}t alors que cette propri\'{e}t\'{e} caract\'{e}ristique des solutions p\'{e}riodiques de type \textit{cycle limite} qui consiste \`{a} se d\'{e}marquer des solutions de type \textit{centre} par leur ind\'{e}pendance vis-\`{a}-vis des conditions initiales \'{e}tait d\'{e}j\`{a} connue et reconnue \`{a} cette \'{e}poque. Pomey rappelle ensuite certains fondements de la th\'{e}orie des \'{e}quations diff\'{e}rentielles lin\'{e}aires avant d'aborder l'\'{e}tude des \textit{oscillations entretenues}. Il explique :

\begin{quote}

\og Pour que des oscillations soient engendr\'{e}es spontan\'{e}ment et s'entretiennent, il ne suffit pas que l'on ait un mouvement p\'{e}riodique, il faut encore que ce mouvement soit stable. \fg{} \cite[p. 375]{Pomey1920}

\end{quote}

De nouveau, Pomey souligne l'importance d'un r\'{e}gime \textit{stable} d'oscillations et prend, pour illustrer son propos, l'exemple de l'oscillateur harmonique (pendule simple) qui poss\`{e}de une solution de type \textit{centre} d\'{e}pendante des conditions initiales. Il d\'{e}montre que dans ce cas : \og on a un r\'{e}gime p\'{e}riodique accidentel, non un r\'{e}gime p\'{e}riodique stable \fg{} \cite[p. 376]{Pomey1920}. Il aborde au {\S 9} l'\'{e}tude d'\textit{\'{e}quations non lin\'{e}aires} et pr\'{e}sente deux exemples pour lesquels la force \'{e}lastique (force de Hooke) n'est plus proportionnelle \`{a} l'\'{e}longation mais s'exprime comme une fonction non lin\'{e}aire (quadratique puis cubique) de l'\'{e}longation. Il fait alors appel \`{a} une technique de d\'{e}veloppement en s\'{e}rie de Fourier, qui n'est pas sans rappeler celle utilis\'{e}e par Blondel \cite{Blondel1919c}, pour expliciter les premiers termes de la solution. Il s'int\'{e}resse ensuite au probl\`{e}me de la \textit{g\'{e}n\'{e}ration des oscillations spontan\'{e}es} et d\'{e}crit au paragraphe {\S 11} un cas particulier qu'il attribue \`{a} Lord Rayleigh et Abraham\footnote{Il semble faire r\'{e}f\'{e}rence d'une part \`{a} l'article de Rayleigh \cite{Rayleigh1883} et, d'autre part aux travaux d'Abraham et Bloch concernant le \textit{multivibrateur} \cite{AbrahamBloch}.}. Pomey propose d'\'{e}tablir l'\'{e}quation diff\'{e}rentielle caract\'{e}risant les oscillations entretenues par un arc chantant aliment\'{e} par une force \'{e}lectromotrice constante continue. Il \'{e}crit :

\begin{quote}

\og Supposons, par exemple, qu'un arc A, aux bornes duquel est branch\'{e} un circuit oscillant, soit aliment\'{e} \`{a} courant constant.

\begin{figure}[htbp]
\centerline{\includegraphics[width=2.82cm,height=3.99cm]{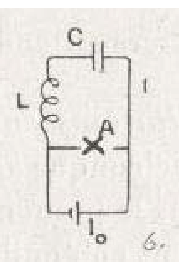}}
\end{figure}

Soit $E$ la diff\'{e}rence de potentiel aux bornes de l'arc, et supposons que cette diff\'{e}rence de potentiel d\'{e}pende du courant oscillant $i$ d'apr\`{e}s la loi

\begin{center}

\hfill $E = E_0 + ai-bi^3$ \hfill (P$_{1}$)

\end{center}

L'\'{e}quation relative du circuit oscillant est, d'autre part,

\begin{center}

\hfill $L \dfrac{di}{dt} + Ri + \dfrac{1}{C} q =E$ \hfill (P$_{2}$)

\end{center}

$q$ \'{e}tant la charge du condensateur \`{a} l'\'{e}poque $t$, quantit\'{e} dont $i$ est la d\'{e}riv\'{e}e par rapport au temps. \fg{} \\ \vphantom{} \hfill \cite[p. 380]{Pomey1920}

\end{quote}

Ainsi, d\`{e}s la fin du mois de juin 1920, Pomey envisageait d\'{e}j\`{a} de repr\'{e}senter la \textit{caract\'{e}ristique d'oscillation} de l'arc chantant, dispositif parfaitement analogue \`{a} la lampe \`{a} trois \'{e}lectrodes d'apr\`{e}s Janet \cite{Janet1919}, par une fonction cubique. Trois semaines plus tard, Van der Pol \cite{VdP1920} employait ce m\^{e}me type de mod\'{e}lisation pour \'{e}tablir l'\'{e}quation diff\'{e}rentielle des oscillations entretenues par une triode. Dans le m\^{e}me paragraphe, Pomey \'{e}tablissait ainsi l'\'{e}quation diff\'{e}rentielle des oscillations entretenues par un arc chantant.

\begin{quote}

\og Des \'{e}quations (P$_{1}$) et (P$_{2}$) on d\'{e}duit, abstraction faite de la constante CE$_0$,

\begin{center}

\hfill $L \dfrac{d^2q}{dt^2} + (R - a)\dfrac{dq}{dt} + \dfrac{1}{C} q = -b \left( \dfrac{dq}{dt} \right)^3 $ \hfill (P$_{3}$).

\end{center}

Consid\'{e}rons donc, \`{a} titre d'exemple d'\'{e}quation non lin\'{e}aire, l'\'{e}quation

\begin{center}

\hfill $L x'' + R x' + \dfrac{1}{C} x = E_0 +a x'- b x'^3 $. \hfill (P$_{4}$).

\end{center}

Posons

\[
\quad x - CE_0 = y \mbox{,} \qquad R - a = \alpha \quad
\]

et remarquons que si l'on fait $R = a$, $b = 0$, l'\'{e}quation a pour solution

\[
\hspace{3cm} y = A \sin \left( mt \right) \mbox{,} \qquad m = \frac{1}{\sqrt{CL}} \mbox{.\fg{} \cite[p. 380]{Pomey1920}}
\]

\end{quote}

En comparant l'\'{e}quation diff\'{e}rentielle (P$_{4}$) avec l'Eq. (HP$_{1}$) (voir p. 8) qu'avait \'{e}tablie Henri Poincar\'{e} \cite[p. 390]{P6} lors des conf\'{e}rences qu'il avait r\'{e}alis\'{e}es en mai-juin 1908 et auxquelles il est possible que Pomey ait assist\'{e}\footnote{Il \'{e}tait pr\'{e}sent lors des ultimes conf\'{e}rences de Poincar\'{e} \cite{P7} en juillet 1912.}, on constate qu'elles sont absolument identiques. Plus extraordinaire encore, Pomey va expliciter les deux premiers termes de la solution de l'\'{e}quation (P$_{4}$) en faisant de nouveau appel \`{a} un d\'{e}veloppement en s\'{e}rie de Fourier :

\begin{quote}

\og On a donc approximativement

\end{quote}

\begin{center}

\hspace{1.2cm} $x - CE_0 = \sqrt{\dfrac{4}{3}\dfrac{(a - R)}{bm^2}}\sin\left(mt \right) + B\cos\left( 3mt \right) + \ldots$ (P$_{4}$). \hspace{1cm}

\end{center}

\begin{quote}

On voit donc qu'il y a une solution p\'{e}riodique ind\'{e}pendante des conditions initiales, qui est une fonction p\'{e}riodique. \fg{} \cite[p. 381]{Pomey1920}

\end{quote}

Cette derni\`{e}re citation implique plusieurs remarques. En tout premier lieu, on constate que Pomey \'{e}tablit de nouveau l'ind\'{e}pendance de la solution p\'{e}riodique vis-\`{a}-vis des conditions initiales (voir Andronov \cite{Andronov1928,Andronov1929}). En second lieu, qu'il fournit les deux premiers termes du d\'{e}veloppement en s\'{e}rie de la solution p\'{e}riodique. Enfin, qu'en faisant abstraction de la r\'{e}sistance $R$, la valeur de l'amplitude de ces oscillations $\sqrt{\dfrac{4}{3}\dfrac{(a - R)}{bm^2}}$ est \'{e}quivalente \`{a} celle obtenue par Blondel (B$_{4}$) pour la lampe \`{a} trois \'{e}lectrodes.

\subsection{Oscillations entretenues par une triode} \hfill

\vspace{0.1in}

En 1916, apr\`{e}s des \'{e}tudes de physique et de math\'{e}matiques \`{a}
l'universit\'{e} d'Utrecht, Balthazar Van der Pol (1889-1959) partit
\'{e}tudier avec John Ambrose Fleming (1849-1945), ing\'{e}nieur \'{e}lectricien anglais qui \'{e}tait professeur \`{a} l'University College de Londres.
Fleming \'{e}tait alors le premier professeur d'\'{e}lectrotechnique de cet
\'{e}tablissement d'enseignement sup\'{e}rieur, mais il \'{e}tait surtout
connu comme l'inventeur de la diode, c'est-\`{a}-dire, la premi\`{e}re valve
thermoïonique, en 1904. Apr\`{e}s une ann\'{e}e avec Fleming, Van der Pol part
travailler avec John Joseph Thomson (1856-1940) au laboratoire Cavendish de Cambridge.
En 1920 il retourna en Hollande pour terminer son doctorat en physique \`{a}
l'universit\'{e} d'Utrecht aux c\^{o}t\'{e}s de Hendrik Lorentz (1853-1928). Sa
th\`{e}se porte sur \og L'influence d'un gaz ionis\'{e} sur la propagation des ondes \'{e}lectromagn\'{e}tiques~: application \`{a} la
t\'{e}l\'{e}graphie sans fil et aux mesures du rayonnement ultra-violet \fg{}.

Ainsi, d\`{e}s le d\'{e}but des ann\'{e}es vingt, Van der Pol s'int\'{e}resse \`{a} la production d'ondes \'{e}lectromagn\'{e}tiques au
moyen de circuits \'{e}lectriques oscillants comportant une triode en lieu et place d'un arc chantant. Cependant, ce n'est pas dans sa c\'{e}l\`{e}bre contribution \og On relaxation-Oscillations \fg{} \cite{VdP1926} mais dans un article ant\'{e}rieur, dont il ach\`{e}ve\footnote{La note de bas de page dans laquelle Van der Pol \cite[p. 702]{VdP1920} fait
r\'{e}f\'{e}rence \`{a} un article de W. E. Eccles publi\'{e} en novembre 1919 qui semble indiquer qu'il n'a commenc\'{e} la r\'{e}daction
qu'apr\`{e}s cette date.} la r\'{e}daction le 17 juillet 1920 et qui est publi\'{e} en novembre et d\'{e}cembre de la m\^{e}me ann\'{e}e, que Van der Pol mod\'{e}lise la caract\'{e}ristique d'oscillation d'une triode au moyen d'une fonction cubique et qu'il \'{e}tablit son \'{e}quation diff\'{e}rentielle, un an apr\`{e}s Blondel \cite{Blondel1919c}. Pour y parvenir, il effectue un d\'{e}veloppement en s\'{e}rie de Taylor-McLaurin de la force \'{e}lectromotrice $\psi \left( {kv} \right)$ de la triode limit\'{e} au trois premiers termes:

\vspace{0.1in}

\begin{center}

\hfill $i = \psi \left( {kv} \right) = - \alpha v + \beta v^2 + \gamma v^3$ \hfill (V$_{1}$)

\end{center}

\vspace{0.1in}

Van der Pol ajoute que, par des consid\'{e}rations de sym\'{e}trie de la caract\'{e}ristique d'oscillation, cette expression peut \^{e}tre r\'{e}duite en posant : $\beta =0$. Deux ans plus tard, pour rendre compte du ph\'{e}nom\`{e}ne d'hyst\'{e}r\'{e}sis d'oscillation de la triode, Appleton et Van der Pol \cite[p. 182]{AppletonVdP1922} seront contraints de d\'{e}velopper la fonction $\psi \left( kv \right)$ jusqu'\`{a} l'ordre cinq, exactement comme l'avait fait Blondel \cite[p. 946]{Blondel1919c}. Il exprime alors la tension aux bornes de chaque dip\^{o}le~:

\[
L\frac{di_1 }{dt} = Ri_3 = \frac{1}{C}\int {i_2 dt} = E_a -v_a
\]

\vspace{0.1in}

et parvient \`{a} \'{e}tablir l'\'{e}quation diff\'{e}rentielle suivante pour la triode :

\[
\frac{di}{dt} + C \frac{d^2v}{dt^2} + \frac{1}{R}\frac{dv}{dt} + \frac{1}{L}v=0
\]

\vspace{0.1in}

En substituant l'expression (V$_{1}$) de l'intensit\'{e} $i$ dans cette \'{e}quation on a~:

\[
C\frac{d^2v}{dt^2} + \left( \frac{1}{R}-\alpha \right) \frac{dv}{dt} + \frac{1}{L}v + \beta \frac{d\left( {v^2} \right)}{dt} + \gamma \frac{d\left( {v^3} \right)}{dt} = 0
\]

\vspace{0.1in}

En posant $\beta = 0$ et en faisant abstraction de la r\'{e}sistance R, ce qui \'{e}quivaut \`{a} $R\to \infty$, cette \'{e}quation devient :

\vspace{0.1in}

\begin{center}

\hfill $C\dfrac{d^2v}{dt^2}-\left( {\alpha -3\gamma v^2} \right)\dfrac{dv}{dt}+\dfrac{1}{L}v=0$ \hfill (V$_{2}$)

\end{center}

\vspace{0.1in}

Pour le calcul de l'amplitude il propose trois m\'{e}thodes. La premi\`{e}re, \og analytique \fg{}, comme l'explique Van der Pol \cite[p. 704]{VdP1920} consiste en un d\'{e}veloppement en perturbations singuli\`{e}res utilis\'{e} par les astronomes. La seconde, faisant appel au d\'{e}veloppement en s\'{e}rie de Fourier utilis\'{e} par Blondel \cite{Blondel1919c} et Pomey \cite{Pomey1920} \'{e}tait plut\^{o}t employ\'{e}e par les ing\'{e}nieurs et le conduit \`{a} introduire une premi\`{e}re correction pour la valeur de la p\'{e}riode plus exactement de la fr\'{e}quence angulaire ou pulsation $\omega$~:

\[
\omega ^2\approx \frac{1}{CL}-\varepsilon \quad \mbox{avec} \quad \varepsilon =\frac{a^2\beta
^2}{3C^2}
\]

\vspace{0.1in}

Mais puisque pour des raisons de sym\'{e}trie $\beta =0$, Van der Pol \cite[p. 705]{VdP1920} obtient~alors :

\vspace{0.1in}

\begin{center}

\hfill $\omega ^2\approx \dfrac{1}{CL}$ \hfill (V$_{3}$)

\end{center}

\vspace{0.1in}

Il est important de souligner que Van der Pol retrouve, comme Blondel, la formule de Thomson en premi\`{e}re approximation. La troisi\`{e}me m\'{e}thode de calcul de l'amplitude est \og g\'{e}om\'{e}trique \fg{} et lui permet comme les deux pr\'{e}c\'{e}dentes d'aboutir \`{a} l'expression suivante :

\[
a=\sqrt {\frac{4}{3}\frac{\alpha -\dfrac{1}{R}}{\gamma }}
\]

En faisant de nouveau abstraction de la r\'{e}sistance $R$ on a finalement :\\

\begin{center}

\hfill $a=\sqrt {\dfrac{4}{3}\dfrac{\alpha}{\gamma }}$ \hfill (V$_{4}$)

\end{center}

\vspace{0.1in}

Il est facile de montrer, comme le fera d'ailleurs Pomey \cite[p. 259]{Pomey1931} que la repr\'{e}sentation de la caract\'{e}ristique d'oscillation propos\'{e}e par Blondel \cite[p. 946]{Blondel1919c} peut se ramener \`{a} celle de Van der Pol \cite[p. 703]{VdP1920}. En effet, en \'{e}crivant l'Eq. (B$_{1}$) ainsi :

\[
i=F\left( u + kv \right) = b_1 \left( 1 + k\frac{v}{u} \right)u -b_3 \left(1 + k\frac{v}{u} \right)^3u^3 + \ldots
\]

et, en utilisant le fait que le potentiel de grille $v \approx - \dfrac{M}{L}u$ on a :

\[
i = b_1 \left( 1 - \frac{kM}{L} \right)u - b_3 \left(1 - \frac{kM}{L} \right)^3 u^3 + \ldots = Au + Bu^3 + \ldots
\]

Ainsi, il appara\^{i}t clairement que, moyennant les simplifications effectu\'{e}es, les \'{e}quations (V$_{1}$, V$_{2}$, V$_{3}$, V$_{4}$) de Van der Pol \cite{VdP1920} et (B$_{1}$, B$_{2}$, B$_{3}$, B$_{4}$) de Blondel \cite{Blondel1919c}
sont totalement identiques au signe pr\`{e}s du sens arbitrairement choisi pour le courant. On remarque \'{e}galement que la mod\'{e}lisation propos\'{e}e par Van der Pol pour la caract\'{e}ristique d'oscillation de la triode est exactement la m\^{e}me que celles initialement introduites par Blondel puis Pomey pour la lampe \`{a} trois \'{e}lectrodes et l'arc chantant.

\newpage

\subsection{Oscillations entretenues par l'arc chantant (II)} \hfill

\vspace{0.1in}

En 1926, Blondel \cite{Blondel1926} publie aux \textit{Comptes-Rendus} une note intitul\'{e}e : \og Contribution \`{a} la th\'{e}orie de l'arc chantant musical \fg{} dans laquelle il \'{e}tablit l'\'{e}quation diff\'{e}rentielle caract\'{e}risant les oscillations du dispositif imagin\'{e} par Duddell \cite{Duddell1900a,Duddell1900b} en reprenant la mod\'{e}lisation employ\'{e}e par Pomey \cite[p. 380]{Pomey1920} sans pour autant faire r\'{e}f\'{e}rence \`{a} ses travaux. Blondel s'int\'{e}resse tout d'abord \`{a} la \textit{courbure de la caract\'{e}ristique} d'oscillation et \'{e}crit :

\begin{quote}

\og Cette caract\'{e}ristique ayant dans sa zone d'utilisation une forme parabolique, nous \'{e}crirons l'expression de la tension aux bornes $u$, en fonction du courant dans l'arc $i$ sous la forme

\[
u = -hi + pi^2 + qi^3 + \ldots.
\]

Nous nous bornerons aux trois premiers termes de la s\'{e}rie, parce que les termes au-del\`{a} du premier n'ont qu'une importance relativement faible. Le premier terme repr\'{e}sente la r\'{e}sistance n\'{e}gative apparente de l'arc, le second tient compte de la courbure parabolique; il est n\'{e}cessaire d'ajouter le troisi\`{e}me pour caract\'{e}riser la dyssim\'{e}trie de cette courbure par rapport \`{a} la verticale passant par le r\'{e}gime moyen. \fg{} \cite[p. 900]{Blondel1926}

\end{quote}

Revenant ensuite sur la notion de \textit{caract\'{e}ristique dynamique} que Blondel \cite{Blondel1905a,Blondel1905b,Blondel1905c} avait introduite en 1905, il ajoute :

\begin{quote}

\og{} On sait que la caract\'{e}ristique dynamique se transforme pendant les oscillations en une courbe ferm\'{e}e de forme analogue \`{a} une ellipse allong\'{e}e. \fg{} \cite[p. 900]{Blondel1926}

\end{quote}

Puis, il consid\`{e}re que l'effet de retard de l'arc (d\'{e}calage entre la tension et l'intensit\'{e}) peut \^{e}tre repr\'{e}sent\'{e} par un terme analogue \`{a} celui qui transcrit l'effet d'une capacit\'{e} et qu'il \'{e}crit :

\[
- \frac{1}{s} \int i dt
\]

\vspace{0.1in}

o\`{u} $s$ d\'{e}signe un coefficient ayant les dimensions d'une capacit\'{e}.
Tenant compte de ce \og retard \fg{} il \'{e}crit la diff\'{e}rence de potentiel aux bornes de l'arc ainsi :

\[
u = -hi + pi^2 + qi^3 + \ldots - \frac{1}{s} \int i dt
\]

\vspace{0.1in}

Il rappelle ensuite l'\og \'{e}quation classique\footnote{D'apr\`{e}s Zenneck \cite[p. 90]{Zenneck} cette \'{e}quation avait \'{e}t\'{e} \'{e}tablie par Gustav Kirchhoff (1824-1887) et Lord Kelvin (William Thomson (1824-1907)).} \fg{} du courant oscillant qui traverse l'arc :

\[
Ri + L\frac{di}{dt} + \frac{1}{C} \int i dt = u
\]

\vspace{0.1in}

En diff\'{e}rentiant par rapport au temps et en remplaçant $u$ par l'expression ci-dessus, il obtient :

\begin{center}

\hfill $L\dfrac{d^2i}{dt^2} + (R -h + 2pi + 3qi^2)\dfrac{di}{dt} + (\dfrac{1}{C} - \dfrac{1}{s})i .$ \hfill (B$_{5}$)

\end{center}

En d\'{e}veloppant dans l'\'{e}quation (B$_{5}$) l'intensit\'{e} $i$ en s\'{e}rie de Fourier et en identifiant terme \`{a} terme, il retrouve l'expression de l'approximation d'ordre un de l'amplitude (B$_{4}$) qu'il avait obtenue quelques ann\'{e}es auparavant ainsi que celle calcul\'{e}e par Pomey \cite[p. 381]{Pomey1920}. Il poursuit les calculs et extrait les approximations d'ordre deux et trois de l'amplitude.\\

L'\'{e}quation (B$_{5}$) \'{e}tablie par Blondel dans cette note qui fut expos\'{e}e le 29 mars 1926 devant les membres de l'Acad\'{e}mie des Sciences est en tous points similaire (hors mis le fait qu'elle ne soit pas d\'{e}dimensionn\'{e}e) \`{a} celle qu'avait pr\'{e}sent\'{e} Van der Pol exactement deux jours auparavant lors d'une conf\'{e}rence expos\'{e}e le 27 mars 1926 devant la \textit{Nederlandsche Natuurkundige Vereeniging}\footnote{Soci\'{e}t\'{e} N\'{e}erlandaise de Physique.} et intitul\'{e}e \og Over Relaxatietrillingen\footnote{\og Sur les oscillations de relaxation~\fg{}.} \fg{}.

\section{Sur les oscillations de relaxation}

Au milieu des ann\'{e}es 1920, Van der Pol \'{e}prouve, de la m\^{e}me mani\`{e}re que Blondel \cite{Blondel1919a}, la n\'{e}cessit\'{e} de clarifier les diff\'{e}rents types d'oscillations. Cependant, ce n'est pas sa c\'{e}l\`{e}bre publication ``On Relaxation-Oscillations'' que Van der Pol introduit ce concept mais dans un court m\'{e}moire publi\'{e} en n\'{e}erlandais le 19 novembre 1925 \cite{VdP1925}. De plus, il est important de remarquer qu'il existe au moins quatre versions diff\'{e}rentes de l'article intitul\'{e} ``On Relaxation-Oscillations'' : deux en n\'{e}erlandais, une en allemand et la derni\`{e}re en anglais r\'{e}dig\'{e}es dans l'ordre chronologique suivant~: \\

- Over Relaxatietrillingen, \textit{Physica}\footnote{La revue \textit{Physica: Nederlands Tijdschrift voor Natuurkunde} fut cr\'{e}\'{e}e en 1921 \`{a} l'initiative d'Adriaan Fokker, d'Ekko Oosterhuis et de Balthazar Van
der Pol. Apr\`{e}s une restructuration en 1934 cette revue a \'{e}t\'{e}
rachet\'{e}e par la soci\'{e}t\'{e} \textit{Elsevier} en 1970 pour prendre la forme connue aujourd'hui sous le nom~de \textit{Physica}.}, 6, p. 154-157,\\

- Over Relaxatie-trillingen,~\textit{Tijdschr. Ned. Radiogenoot.} 3, p. 25-40,\\

- \"{U}ber Relaxationsschwingungen, \textit{Jb. Drahtl. Telegr.} 28, p. 178-184,\\

- On relaxation-oscillations, \textit{Philosophical Magazine}, 7, 2 p. 978-992.

\subsection{La g\'{e}n\'{e}ricit\'{e} de l'\'{e}quation de Van der Pol}\hfill

\vspace{0.1in}

Dans ses publications sur les oscillations de relaxation, Van der Pol pr\'{e}sente une \'{e}quation diff\'{e}rentielle qui n'est plus rattach\'{e}e \`{a} la triode, ni \`{a} aucun autre dispositif (machine s\'{e}rie-dynamo ou arc chantant). Avec une
d\'{e}marche tr\`{e}s p\'{e}dagogique qui n'est pas sans rappeler celle de Pomey \cite[p. 77]{Pomey1920} il pr\'{e}sente l'\'{e}quation de
l'oscillateur amorti et le concept de \og r\'{e}sistance n\'{e}gative \fg{} en consid\'{e}rant que si le signe du coefficient de frottements $\alpha$ est invers\'{e}, ce qui se produit notamment dans la machine s\'{e}rie-dynamo, l'arc chantant et la triode, alors l'amplitude des oscillations augmente ind\'{e}finiment ce qui est physiquement irr\'{e}alisable. Il explique ensuite que pour limiter l'amplitude on doit supposer que le coefficient de frottements est une fonction de l'amplitude. Il propose donc de remplacer $\alpha$ par l'expression $\alpha -3\gamma x^2$ (o\`{u} $\gamma $ est une constante) dans l'\'{e}quation de l'oscillateur amorti\footnote{Ce type de mod\'{e}lisation au moyen d'une focntion cubique avait d\'{e}j\`{a} \'{e}t\'{e} envisag\'{e}e en 1883 par Lord Rayleigh \cite[p. 230]{Rayleigh1883}.}. Il obtient l'\'{e}quation suivante~:

\[
\ddot {x}-\left( {\alpha -3\gamma x^2} \right)\dot {x}+\omega^2x=0
\]

\vspace{0.1in}

Contrairement aux apparences, cette \'{e}quation n'est pas sans dimensions et correspond \`{a} l'\'{e}quation (V$_{2}$) repr\'{e}sentant les oscillations d'une triode pr\'{e}c\'{e}demment \'{e}tablie par Van der Pol \cite{VdP1920}. La variable $x$ a donc la dimension d'une diff\'{e}rence de potentiel (tension de plaque de la triode). Sous cette forme l'\'{e}quation poss\`{e}de trois param\`{e}tres : $\alpha$ la \og r\'{e}sistance n\'{e}gative \fg{}, $\omega$ la fr\'{e}quence angulaire ou pulsation d\'{e}finie par (V$_{3}$) et $\gamma$ une constante. Van der Pol fait alors appel \`{a} une technique introduite par Pierre Curie \cite{Curie} sans cependant y faire r\'{e}f\'{e}rence. Il obtient ainsi une \'{e}quation diff\'{e}rentielle sans dimensions et qui ne d\'{e}pend plus d\'{e}sormais que d'un seul param\`{e}tre~:\\

\begin{center}

\hfill $\ddot {v}-\varepsilon \left( {1-v^2} \right)\dot {v}+v=0$ \hfill (V$_{5}$)

\end{center}

\vspace{0.1in}

Cette forme r\'{e}duite (V$_{5}$) est appel\'{e}e \og \'{e}quation de Van der Pol\footnote{Une correspondance biunivoque entre cette \'{e}quation et celle \'{e}tablie par Lord Rayleigh \cite{Rayleigh1883} a \'{e}t\'{e} mise en \'{e}vidence par Le Corbeiller \cite[p. 723]{LeCorbeiller1932}. Voir Ginoux \cite{Gith}.}~\fg{}.

\subsection{L'int\'{e}gration graphique et les oscillations de relaxation}\hfill

\vspace{0.1in}

Tr\`{e}s t\^{o}t, Van der Pol avait pris conscience du caract\`{e}re non-int\'{e}grable de l'\'{e}quation (V$_{5}$) qu'il rappelle
ici au tout d\'{e}but de son article~:

\begin{quote}

\og It has not been found possible to obtain an approximate analytical solution for (V$_{5}$) with the supplementing condition ($\varepsilon \gg 1$), but a graphical solution may be found in the following way. \fg{} \cite[p. 982]{VdP1926}

\end{quote}

L'int\'{e}gration graphique de l'\'{e}quation (V$_{5}$) va conduire Van der Pol [1926c, p. 987] \`{a} \'{e}tablir que la p\'{e}riode
des oscillations dans le troisi\`{e}me cas, $i.e.$ pour $\varepsilon =10$ devient approximativement \'{e}gale \`{a} $\varepsilon $.

\begin{figure}[htbp]
\centerline{\includegraphics[width=13.825cm,height=8.1855cm]{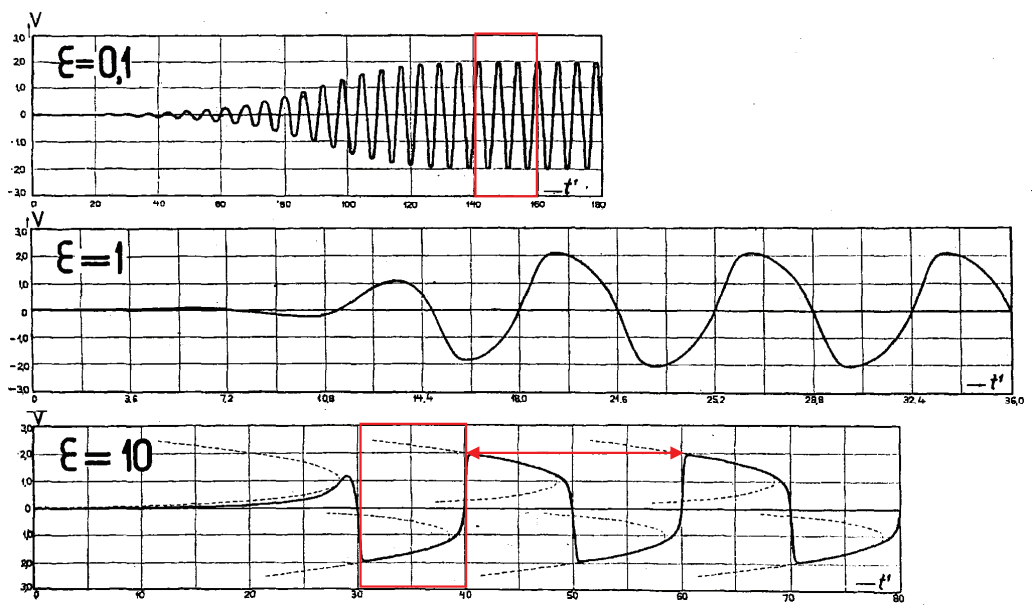}}
\caption[Int\'{e}gration graphique de l'\'{e}quation de Van der Pol]{Int\'{e}gration graphique de l'\'{e}quation (V$_{5}$), \\ \centering{d'apr\`{e}s Van der Pol \cite[p. 986]{VdP1926}.}}
\label{fig11}
\end{figure}

Sur la premi\`{e}re courbe ($\varepsilon =0.1)$ on d\'{e}nombre, dans la partie encadr\'{e}e en rouge, trois p\'{e}riodes un quart pour vingt graduations temporelles. Ce qui fournit une p\'{e}riode T approximativement
\'{e}gale \`{a} $2\pi $ comme l'affirme Van der Pol \cite[p. 987]{VdP1926}.\\

La seconde courbe ($\varepsilon =1)$ permet de mettre en lumi\`{e}re \og une d\'{e}viation assez sensible de la forme sinusoïdale. \fg{} \cite[p. 17]{VdP1930}. Elle montre la transition continue entre les oscillations
sinusoïdales ($\varepsilon \ll 1$) et les oscillations de relaxation ($\varepsilon \gg 1$).\\

Dans la partie encadr\'{e}e en rouge de la troisi\`{e}me courbe ($\varepsilon =10)$ on constate que \og la courbe commence \`{a} s'\'{e}lever exponentiellement, et au bout d'une dur\'{e}e \'{e}gale \`{a} une p\'{e}riode, a d\'{e}j\`{a} pratiquement atteint sa forme p\'{e}riodique limite. \fg{}  \cite[p. 18]{VdP1930}. Ainsi, il en d\'{e}duit que pour $\varepsilon \gg 1$ la valeur de la p\'{e}riode est approximativement \'{e}gale \`{a} $\varepsilon $. Mais, puisque $\varepsilon$ est un param\`{e}tre sans dimension il lui faut donc exprimer la p\'{e}riode dans un syst\`{e}me d'unit\'{e}s temporelles. En appliquant la formule de changement de variable, ${t}'=\omega t$ la p\'{e}riode s'\'{e}crit :

\[
T=\frac{\varepsilon }{\omega }
\]

En tenant compte du fait qu'il a pos\'{e} : $\varepsilon =\dfrac{\alpha }{\omega }$,
$\alpha =\dfrac{R}{L}$ et $\omega ^2=\dfrac{1}{LC}$ il obtient~finalement :

\begin{center}

\hfill $T=\dfrac{\alpha }{\omega ^2}=RC$ \hfill (V$_{6}$)

\end{center}

\vspace{0.1in}

Il d\'{e}montre ainsi que pour des valeurs de $\varepsilon >>1$ la p\'{e}riode des oscillations est approximativement\footnote{N\'{e}anmoins, il sera \'{e}tabli ensuite par Li\'{e}nard \cite[p. 952]{Liénard1928} que la p\'{e}riode vaut en r\'{e}alit\'{e} le double soit $2\varepsilon$ comme il est facile de le v\'{e}rifier sur la Fig. 11.} \'{e}gale au \og temps de relaxation \fg{} correspondant \`{a} la dur\'{e}e de d\'{e}charge d'un condensateur dans une r\'{e}sistance et propose alors d'appeler ce ph\'{e}nom\`{e}ne \og oscillations de relaxation \fg{}. Van der Pol rappelle ensuite que cette id\'{e}e lui a \'{e}t\'{e} sugg\'{e}r\'{e}e par la lecture d'un article d'Abraham et Bloch \cite{AbrahamBloch}  dans lequel ils d\'{e}crivent le fonctionnement du \textit{multivibrateur}, si\`{e}ge d'oscillations dont la p\'{e}riode est \og de l'ordre de $C_1 R_1 +C_2 R_2 $ \fg{}. Van der Pol explique~:

\begin{quote}

\og In their original description of the system Abraham {\&} Bloch draw
attention to the fact that the time period of the oscillations produced by
the multivibrator~is approximately equal to the product RC, but, so, as far
as I am aware, no theoretical discussion of the way in which the
oscillations are \textit{maintained} has been published.  \fg{} \\ \vphantom{} \hfill \cite[p. 988]{VdP1926}

\end{quote}

Ceci para\^{i}t confirmer que ce sont bien les travaux d'Abraham et Bloch qui ont conduit Van der Pol \`{a} expliciter la p\'{e}riode d'oscillations de la triode comme le produit \textit{RC}. De plus, il semble que ce soit \'{e}galement leur description du ph\'{e}nom\`{e}ne, de charge \og brusque \fg{}  et de d\'{e}charge \og lente \fg{}, qui l'ait incit\'{e} \`{a}
donner \`{a} ces oscillations qu'il qualifiait encore quelques ann\'{e}es auparavant de \textit{maintained} le nom d'oscillations de relaxation\footnote{Dans son analyse de l'article de Van der Pol \cite{VdP1926}, l'ing\'{e}nieur Pierre David (1897-1987) proposait de traduire cette terminologie par \og oscillations par d\'{e}charge \fg{} \cite{David} qui n'est pas sans rappeler l'analogie avec la d\'{e}charge fractionn\'{e}e de Gaugain mise en \'{e}vidence par Blondel \cite{Blondel1905a,Blondel1905b}.}.

Ainsi, le grand m\'{e}rite de Van der Pol est d'avoir d'une part mis en lumi\`{e}re le caract\`{e}re \og lent-rapide \fg{} de ces oscillations et, d'autre part d'avoir d\'{e}montr\'{e} que d'autres dispositifs sont r\'{e}gis par l'\'{e}quation d\'{e}dimensionn\'{e}e (V$_{5}$) et sont donc le si\`{e}ge d'oscillations de relaxation. L'ing\'{e}nieur Philippe Le Corbeiller (1891-1980) qui assista Van der Pol lors de ses conf\'{e}rences \`{a} Paris de 1928 \`{a} 1930 et qui participa grandement \`{a} la diffusion en France du concept d'oscillations de relaxation rappelait lors de son expos\'{e} au Conservatoire des Arts et M\'{e}tiers l'importance de la contribution de Van der Pol en ces termes :

\begin{quote}

\og \textit{Cas de} $\varepsilon $ \textit{tr\`{e}s grand}. -- Ici la courbe des oscillations a visiblement un tr\`{e}s grand nombre d'harmoniques~; en langage math\'{e}matique, la s\'{e}rie de Fourier correspondante converge tr\`{e}s lentement. Il est donc absolument illusoire dans ce cas de calculer plus ou moins p\'{e}niblement les un, deux ou trois premiers termes de la s\'{e}rie. Un des apports les plus consistants de M. van der Pol a consist\'{e} \`{a} reconna\^{i}tre clairement ce fait, \`{a} donner \textit{un nom} \`{a} ces oscillations non sinusoïdales, et \`{a} en faire un outil de la recherche physique, au m\^{e}me titre que les oscillations sinusoïdales dont elles constituent en fait la contrepartie. \fg{} \cite[p. 22]{LeCorbeiller1931}

\end{quote}

\newpage

\section{La correspondance Blondel-Cartan} \hfill

Une lettre adress\'{e}e par Andr\'{e} Blondel au math\'{e}maticien \'{E}lie Cartan vient r\'{e}cemment d'\^{e}tre d\'{e}couverte\footnote{Ce document m'a \'{e}t\'{e} aimablement transmis par M. Scott Walter.}. Dans ce courrier dat\'{e} du 8 avril 1931, Blondel remercie tout d'abord Cartan pour les tir\'{e}s \`{a} part de son article \cite{Cartan} puis rend hommage aux \og grands math\'{e}maticiens \fg{} qu'il consid\`{e}re comme \og les ma\^{i}tres de la science \fg{} susceptibles d'apporter des solutions aux probl\`{e}mes pos\'{e}s par d'autres sciences comme la M\'{e}canique ou la Physique. Il \'{e}crit :

\begin{quote}

\og Vous venez d'en donner un nouvel exemple par votre intervention dans la th\'{e}orie des oscillations de relaxation (que j'ai apprise il y a 12 ans ``oscillations de 2\`{e}me esp\`{e}ce'') et dans la question si \'{e}pineuse de la relativit\'{e} g\'{e}n\'{e}ralis\'{e}e, gr\^{a}ce \`{a} vous Einstein a trouv\'{e} en France \`{a} qui parler ; c'est un grand soulagement pour nous en m\^{e}me temps qu'une v\'{e}ritable satisfaction pour tous ceux qui souhaitent la mise au point d\'{e}finitive de cette branche si complexe de nos connaissances. \fg{}

\end{quote}

Dans la premi\`{e}re phrase, Blondel rappelle que douze ans auparavant il avait mis en \'{e}vidence un nouveau type d'oscillations qu'il avait alors qualifi\'{e} de \og seconde esp\`{e}ce \fg{}. En effet, dans cet article Blondel \cite{Blondel1919a} avait d\'{e}fini dans le second type (2.) des \og oscillations persistantes par \'{e}coulement fractionn\'{e} \fg{}. Il avait ensuite illustr\'{e} ce type d'oscillations \`{a} partir de \og l'arc de seconde esp\`{e}ce ou arc sifflant \fg{} qu'il avait d\'{e}couvert en 1905 (voir Blondel \cite{Blondel1905a,Blondel1905b,Blondel1905c}).\\

\begin{figure}[htbp]
\centerline{\includegraphics[width=9.243cm,height=7.857cm]{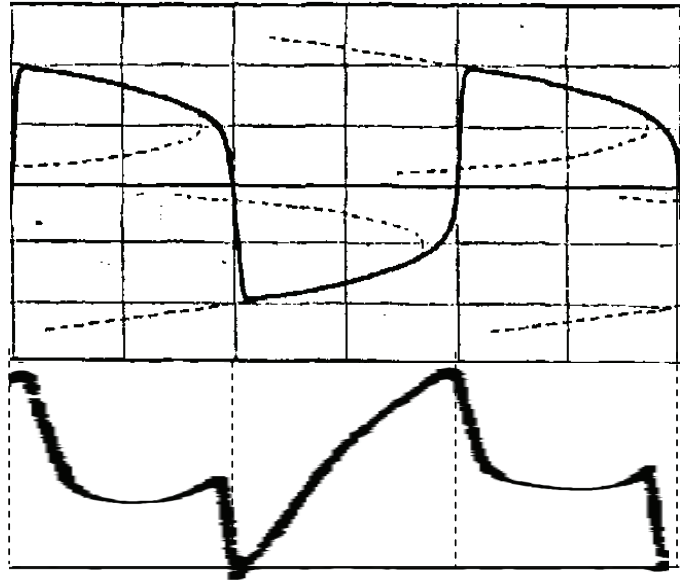}}
\caption[Oscillations de relaxation dans l'arc sifflant]{Oscillations de relaxation dans l'arc sifflant,\\ \centering{d'apr\`{e}s Van der Pol \cite[p. 986]{VdP1926} et Blondel \cite[p. 78]{Blondel1905b}}.}
\label{fig12}
\end{figure}

Sur la Fig. 12 ont \'{e}t\'{e} repr\'{e}sent\'{e}es les oscillogrammes, c'est-\`{a}-dire, les s\'{e}ries temporelles de l'\'{e}volution de la solution p\'{e}riodique correspondant d'une part au ph\'{e}nom\`{e}ne d'\textit{oscillations de relaxation} mis en \'{e}vidence par Van der Pol \cite{VdP1925, VdP1926} (partie sup\'{e}rieure) et, d'autre part au \textit{second type discontinu} d'oscillations d\'{e}couvert par Blondel \cite{Blondel1905a,Blondel1905b,Blondel1905c} (partie inf\'{e}rieure\footnote{La Fig. 4 (voir p. 11) a \'{e}t\'{e} d\'{e}form\'{e}e afin de permettre une comparaison.}). On constate une parfaite analogie dans la forme de ces oscillations que Van der Pol ne tardera pas \`{a} faire remarquer.

En effet, le 11 mars 1930, lors de sa seconde conf\'{e}rence \`{a} l'\'{E}cole Sup\'{e}rieure d'\'{E}lectricit\'{e}, Van der Pol \cite{VdP1930} reprenait son \'{e}num\'{e}ration des diff\'{e}rents dispositifs si\`{e}ges d'\textit{oscillations de relaxation} :

\begin{quote}

\og Dans le domaine \'{e}lectrique nous avons de tr\`{e}s jolis exemples d'oscillations de relaxation dont certains sont tr\`{e}s anciens, telles que la d\'{e}charge par \'{e}tincelle d'une machine \`{a} plateaux, l'oscillation de l'arc \'{e}lectrique \'{e}tudi\'{e}e par M. {\sc Blondel}, dans un m\'{e}moire c\'{e}l\`{e}bre ($^{1}$) ou l'exp\'{e}rience de M. {\sc Janet}, et d'autres plus r\'{e}centes telles que le multivibrateur d'{\sc Abraham} et {\sc Bloch} ($^{2}$) qui consiste en un amplificateur \`{a} couplage capacit\'{e} et
r\'{e}sistance \`{a} deux \'{e}tages, dont les bornes de sortie sont connect\'{e}es aux bornes d'entr\'{e}e.

\noindent\hrulefill

\scriptsize{($^{1}$) {\sc Blondel}, \textit{Eclair. Elec.}, \textbf{44}, 41, 81, 1905. V. aussi \textit{J. de Phys.}, \textbf{8}, 153, 1919.

($^{2}$) {\sc Abraham} et {\sc Bloch}, \textit{Ann. de Phys.}, \textbf{12}, 237, 1919.} \normalsize{ \fg{} \\ \vphantom{} \hfill \cite[p. 20]{VdP1930}}

\end{quote}

C'est durant cet expos\'{e} (auquel a peut-\^{e}tre assist\'{e} Blondel) que Van der Pol faisait la seule et unique r\'{e}f\'{e}rence aux travaux de Blondel \cite{Blondel1905a,Blondel1905b,Blondel1905c} concernant l'arc \'{e}lectrique. Il est important de remarquer que moins d'un an s'est \'{e}coul\'{e} entre la conf\'{e}rence de Van der Pol et la lettre de Blondel.

\newpage

\section{Conclusion}\hfill

Au regard de ce qui vient d'\^{e}tre expos\'{e} il appara\^{i}t que dans une s\'{e}rie d'articles \og Sur les ph\'{e}nom\`{e}nes de l'arc chantant \fg{} de 1905, Blondel \cite{Blondel1905a,Blondel1905b,Blondel1905c} a mis en \'{e}vidence dans l'\textit{arc sifflant} ce qu'il nomme un \textit{second type discontinu} d'oscillations et qui s'av\`{e}re \^{e}tre des \textit{oscillations de relaxation}. Pour expliquer ce ph\'{e}nom\`{e}ne il propose de repr\'{e}senter son \'{e}volution dans le \textit{plan de phase} et montre que les trajectoires prennent alors l'allure de petits cycles. Cependant, la non reproductibilit\'{e} des exp\'{e}riences \`{a} l'identique du fait de la grande variabilit\'{e} des diff\'{e}rents \'{e}l\'{e}ments constitutifs de l'arc chantant (diam\`{e}tre, \'{e}cart et nature des charbons de l'arc) rendait alors impossible de v\'{e}rifier le caract\`{e}re attractif de ces cycles.\\

Le d\'{e}veloppement consid\'{e}rable que connut la lampe \`{a} trois \'{e}lectrodes pendant le premier conflit mondial conduit Blondel \`{a} transposer les r\'{e}sultats qu'il avait obtenus pour l'arc chantant \`{a} ce nouveau dispositif. Plus fiable et moins instable, la triode allait ainsi faciliter l'\'{e}tude des oscillations entretenues. En avril 1919, Blondel \cite{Blondel1919a} publie un long m\'{e}moire dans lequel il propose une nouvelle classification des oscillations et introduit la terminologie \textit{oscillations auto-entretenues}. Apr\`{e}s avoir donn\'{e} une d\'{e}finition de ce concept qui sera ensuite reprise par Andronov \cite{Andronov1929} par Van der Pol \cite{VdP1930} et par Le Corbeiller \cite{LeCorbeiller1931}, Blondel emploie un diagramme permettant de repr\'{e}senter l'amplitude des oscillations pour illustrer les oscillations (\textit{auto})-\textit{entretenues} par une lampe \`{a} trois \`{a} \'{e}lectrodes et par un arc chantant. Si les figures qu'il obtient semblent parfaitement correspondre aux cycle limites de Poincar\'{e} \cite[p. 261]{P2} il ne fait cependant aucun lien avec les travaux de \og l'illustre g\'{e}om\`{e}tre \fg{} qu'il semble portant bien conna\^{i}tre. De plus, alors que les \og cycles de Blondel \fg{} permettent d'expliquer le m\'{e}canisme de l'entretien par la pr\'{e}sence dans chacun de ces dispositifs d'un composant \`{a} caract\'{e}ristique d'oscillation non lin\'{e}aire, c'est-\`{a}-dire, pour lequel la valeur de la r\'{e}sistance peut pendant un certain laps de temps devenir n\'{e}gative, ils ne fournissent en revanche aucune indication concernant la forme des oscillations que Janet puis Blondel consid\`{e}rent d\'{e}sormais comme non-sinusoïdales. En effet, il faudra attendre les travaux de Van der Pol \cite{VdP1926} pour que l'on s'int\'{e}resse \`{a} la nature intrins\`{e}que de ces oscillations auxquelles il donnera le nom d'\textit{oscillations de relaxation}.\\

\newpage

Entre novembre 1919 et juin 1920, Blondel \cite{Blondel1919c} puis Pomey \cite{Pomey1920} \'{e}tablissent l'\'{e}quation diff\'{e}rentielle des oscillations entretenues par une lampe \`{a} trois \'{e}lectrodes et par un arc chantant en mod\'{e}lisant la caract\'{e}ristique d'oscillation de chacun de ces dispositifs au moyen d'une fonction \textit{quintique} et respectivement d'une \textit{cubique}.\\

En mars 1926, en empruntant \`{a} Pomey \cite{Pomey1920} la repr\'{e}sentation par une fonction \textit{cubique} de sa caract\'{e}ristique d'oscillation, Blondel \cite{Blondel1926} obtient l'\'{e}quation diff\'{e}rentielle des oscillations entretenues par l'arc chantant de Duddell \cite{Duddell1900a,Duddell1900b} qui est en tous points similaire \`{a} celle qu'\'{e}tablit concomitamment Van der Pol \cite{VdP1926}. Ce fait est particuli\`{e}rement troublant et on peut naturellement s'interroger sur la connaissance que Van der Pol a pu avoir des travaux de Blondel et inversement. Cependant, il para\^{i}t tr\`{e}s difficile de r\'{e}pondre \`{a} cette question. Il faudrait pour cela avoir acc\`{e}s aux archives personnelles de Van der Pol ainsi qu'\`{a} celles de Blondel afin de rechercher une trace mat\'{e}rielle comme par exemple un des articles de Blondel \cite{Blondel1905a,Blondel1905b,Blondel1905c, Blondel1919a,Blondel1919b,Blondel1919c,Blondel1919d,Blondel1920,Blondel1926} qui aurait pu \^{e}tre en possession de Van der Pol et \textit{vice versa} ou une correspondance dans laquelle Blondel aurait cit\'{e} les travaux de Van der Pol \cite{VdP1920,VdP1926}. On peut n\'{e}anmoins faire remarquer que Blondel qui \'{e}tait \^{a}g\'{e} en 1926 de soixante-trois ans faisait autorit\'{e} dans le domaine de l'\'{E}lectrotechnique et de la Radio-\'{e}lectricit\'{e}. Laur\'{e}at du Franklin Institute en 1912, il avait \'{e}t\'{e} \'{e}lu membre de l'Acad\'{e}mie des Sciences en 1913 et, avait reçu par la suite de nombreuses distinctions honorifiques comme la M\'{e}daille Mascart ou le prix Montefiore. Il avait \'{e}t\'{e} \'{e}galement Pr\'{e}sident d'honneur de la Soci\'{e}t\'{e} française des \'{E}lectriciens, Vice-Pr\'{e}sident de \textit{The Illuminating Engineering Society} de Londres et Vice-Pr\'{e}sident de la Soci\'{e}t\'{e} Internationale de \'{E}lectriciens. Il avait \'{e}t\'{e} membre du Comit\'{e} Scientifique de la revue \textit{La Lumi\`{e}re \'{E}lectrique} aux c\^{o}t\'{e}s d'Henri Poincar\'{e}, de la revue allemande \textit{Jahrbuch der drahtlosen Telegraphic und Telephonie} aux c\^{o}t\'{e}s notamment de John Ambrose Fleming aupr\`{e}s duquel travailla Van der Pol et Pr\'{e}sident du comit\'{e} de r\'{e}daction de la \textit{Revue G\'{e}n\'{e}rale de l'\'{E}lectricit\'{e}}. Ses travaux avaient connu une notori\'{e}t\'{e} internationale. Aussi, il est tr\`{e}s peu probable que Van der Pol n'en n'ait pas eu connaissance. En ce qui concerne, le math\'{e}maticien russe Aleksandr' Andronov pour lequel le m\^{e}me type d'interrogation peut se poser, on trouve dans l'\'{e}dition originale russe de son ouvrage {\cyr Teoriya kolebani{\u i}} (\textit{Th\'{e}orie des oscillations}) \cite[p. 217, p. 252, p. 279, p. 412]{Andronov1937} plusieurs paragraphes d\'{e}di\'{e}s aux oscillations entretenues par un arc \'{e}lectrique qui renvoient aux travaux de Blondel \cite{Blondel1919a,Blondel1919c} et montrent qu'ils ont servis de pr\'{e}ablable \`{a} Andronov pour l'\'{e}tude de dispositifs plus complexes comme le \textit{multivibrateur} d'Abraham et Bloch par exemple.\\

Ainsi, si Blondel para\^{i}t avoir devanc\'{e} Van der Pol d'une ann\'{e}e dans dans la mise en \'{e}quation compl\`{e}te de la lampe \`{a} trois \'{e}lectrodes et dans la d\'{e}finition d'un \textit{syst\`{e}me auto-entretenu}, il ne semble pas avoir \'{e}t\'{e} en mesure de mettre en \'{e}vidence le caract\`{e}re \textit{lent-rapide} propre aux \textit{oscillations de relaxation} alors qu'il fut pourtant l'un des premiers a les observer dans l'\textit{arc chantant}. N\'{e}anmoins, sa contribution \`{a} l'\'{e}mergence des concepts d'\textit{oscillations de relaxation} par Van der Pol et d'\textit{oscillations auto-entretenues} par Andronov para\^{i}t avoir \'{e}t\'{e} un \'{e}l\'{e}ment indispensable \`{a} l'\'{e}laboration de la th\'{e}orie des oscillations non lin\'{e}aires.

\section*{Acknowledgments}

Je souhaite adresser tous mes remerciements \`{a} M. Scott Walter qui m'a transmis cette correspondance entre A. Blondel et E. Cartan.

\newpage

\section*{Appendix}

Blondel a sans doute observ\'{e} des oscillations entretenues lors
d'exp\'{e}rimentations sur son oscilloscope comme sur la Fig. 13, et pour
expliquer leur construction (Figs. 8, 9 et 10), il va les comparer \`{a} une
sinusoïde de r\'{e}f\'{e}rence (en pointill\'{e}s sur la figure).

\begin{figure}[htbp]
\centerline{\includegraphics[width=4.73in,height=2.92in]{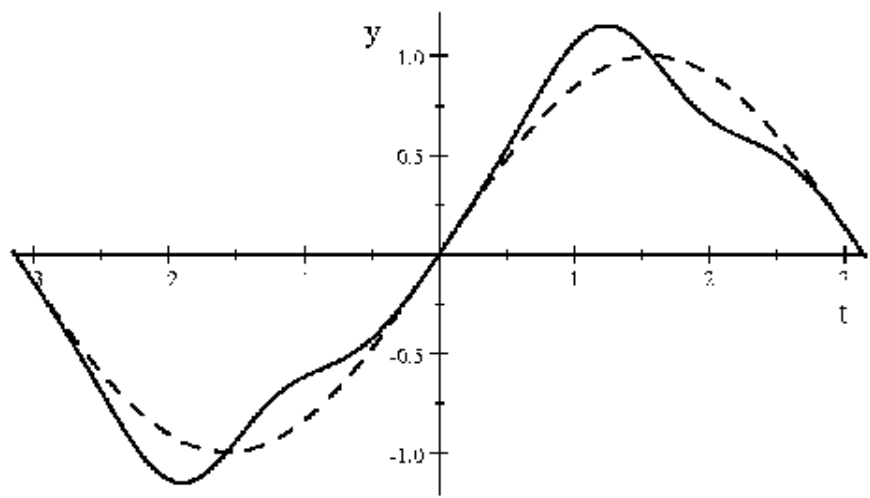}}
\caption[Ligne continue~: une p\'{e}riode d'oscillation entretenue]{Ligne continue~: une p\'{e}riode d'oscillation entretenue, ligne pointill\'{e}e : fonction sinus.}
\label{fig13}
\end{figure}

Il pr\'{e}cise cette diff\'{e}rence en utilisant la formule

\begin{equation}
\rho =Ae^{(-\alpha +j\beta )t}
\label{eq1}
\end{equation}

qui lui permet de tracer un diagramme comme \`{a} la Fig. 14 (o\`{u} $A=1)$
que l'on peut comparer \`{a} la Fig. 9.

\begin{figure}[htbp]
\centerline{\includegraphics[width=2.92in,height=2.92in]{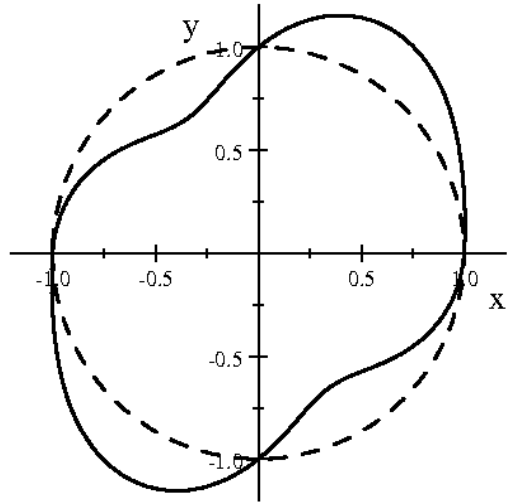}}
\caption[Ligne ferm\'{e}e continue~: une p\'{e}riode d'oscillation
entretenue]{Ligne ferm\'{e}e continue~: une p\'{e}riode d'oscillation
entretenue, ligne pointill\'{e}e~: fonction sinus repr\'{e}sent\'{e}e sous forme d'un cercle trigonom\'{e}trique.}
\label{fig14}
\end{figure}

Cette repr\'{e}sentation transforme la sinusoïde de r\'{e}f\'{e}rence
en un cercle qui va lui servir de gabarit pour appuyer sa construction
d'oscillation entretenue~; son but \'{e}tant de construire une courbe
ferm\'{e}e, form\'{e}e en partie de ce qu'il appelle spirales logarithmiques
(aujourd'hui on dirait plut\^{o}t exponentielles), qui oscille autour du
cercle de r\'{e}f\'{e}rence. Cette repr\'{e}sentation peut toutefois gagner
encore en clart\'{e} si le graphique de la fonction (\ref{eq1}) n'est pas trac\'{e}
dans le syst\`{e}me de coordonn\'{e}es cart\'{e}sienne $(x,y)$ comme \`{a}
la Fig. 14, mais plut\^{o}t dans le syst\`{e}me de coordonn\'{e}es $(t,\rho
)$ qui repr\'{e}sente le temps sur l'axe des abscisses et le rayon sur l'axe
des ordonn\'{e}es (Fig. 15), la sinusoïde de la Fig. 13 \'{e}tant alors
repr\'{e}sent\'{e}e par la droite d'ordonn\'{e}e $\rho =A=1$.

\begin{figure}[htbp]
\centerline{\includegraphics[width=4.73in,height=2.92in]{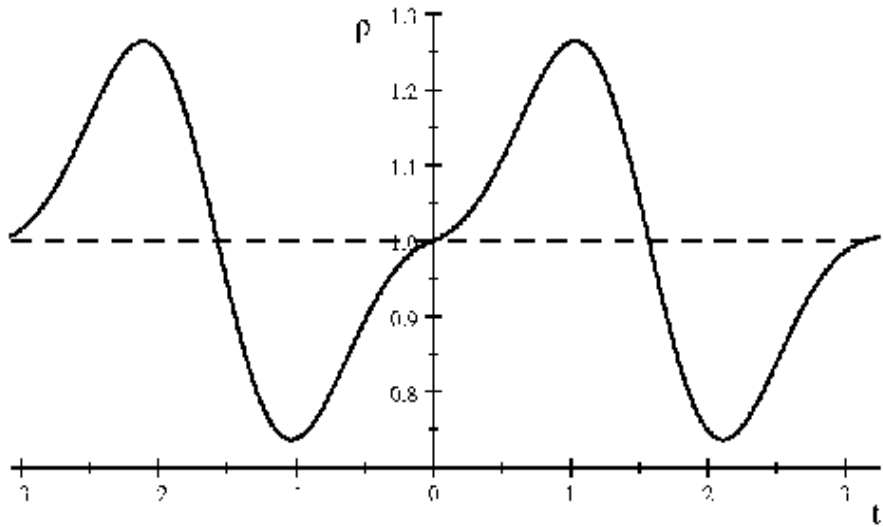}}
\caption[Ligne continue~: une p\'{e}riode d'oscillation entretenue]{Ligne continue~: une p\'{e}riode d'oscillation entretenue, ligne pointill\'{e}e~: fonction sinus repr\'{e}sent\'{e}e sous forme d'une droite, $A=1$.}
\label{fig15}
\end{figure}

Ces nouvelles coordonn\'{e}es permettent de d\'{e}velopper sans limite les
courbes lorsqu'elles tournent autour du centre O, au lieu de les concentrer
sur une partie born\'{e}e du plan comme dans les figures de Blondel. Elles
permettent aussi de mieux observer la convergence de plusieurs courbes
issues de conditions initiales diff\'{e}rentes qui tendraient vers un cycle
limite, sans qu'elles ne se superposent comme elles le feraient sur la Fig.
8. L'\'{e}cart entre l'oscillation entretenue et la sinusoïde (ou le
cercle) de r\'{e}f\'{e}rence \'{e}tant repr\'{e}sent\'{e} plus
explicitement~: sur la Fig. 15, c'est l'\'{e}cart vertical entre la courbe
continue et la droite en pointill\'{e}s.

Pour r\'{e}examiner les figures de Blondel dans ce nouveau syst\`{e}me de
coordonn\'{e}es, on peut r\'{e}\'{e}crire l'\'{e}quation (\ref{eq1}) sous la forme

\begin{equation}
\begin{aligned}
z(t) = x(t) + iy(t) & = r(t)\cos (\beta t)+ir(t)\sin (\beta t) \\
& = Ae^{-\alpha t}\cos(\beta t)+iAe^{-\alpha t}\sin (\beta t)
\end{aligned}
\label{eq2}
\end{equation}

La Fig. 8 se transforme alors en la Fig. 16.

\begin{figure}[htbp]
\centerline{\includegraphics[width=4.73in,height=2.92in]{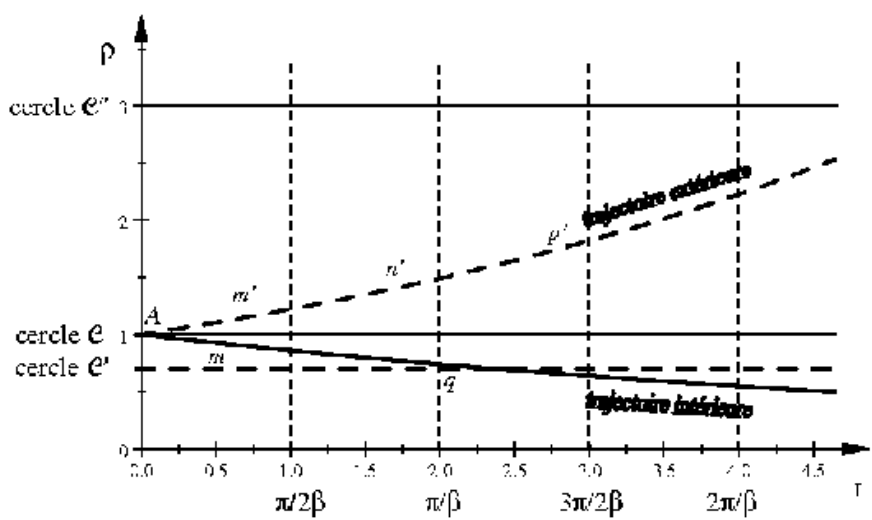}}
\caption[Repr\'{e}sentation de la Fig. 8]{Repr\'{e}sentation de la Fig. 8 dans le syst\`{e}me de
coordonn\'{e}es $(t,\rho )$, $\beta =\frac{\pi }{2}$.}
\label{fig16}
\end{figure}

Les fonctions $\sin$ et $\cos$ de l'\'{e}quation (\ref{eq2}) n'ont pas \`{a} \^{e}tre repr\'{e}sent\'{e}es, ce
qui en simplifie l'analyse. Le centre O de la Fig. 8 est d\'{e}velopp\'{e} en la droite de l'axe du temps sur la Fig. 16. Les trois cercles C, C', C''
sont repr\'{e}sent\'{e}s par des droites d'ordonn\'{e}e respectives~: $\rho=A$, $\rho =C'$, $\rho =C''$. La spirale logarithmique de Blondel
d\'{e}crite par le point mobile $m$, devient une courbe homoth\'{e}tique \`{a}
une fonction exponentielle $r(t)=Ae^{-\alpha t}$, qui partant du point
$(O,\;A)$ va d\'{e}cro\^{i}tre et traverser le cercle C' pour $\alpha >0$,
c'est-\`{a}-dire $-\alpha <0$, ou cro\^{\i}tre et traverser le cercle C''
pour $\alpha >0$ mais quand $-\alpha $ est remplac\'{e} par $+\alpha $ comme
indiqu\'{e}~:

\begin{quote}

\og{}{\ldots} Pour qu'il y ait stabilit\'{e}, il est n\'{e}cessaire que le
ph\'{e}nom\`{e}ne d'entretien donne au syst\`{e}me oscillant un
amortissement tr\`{e}s petit, mais l\'{e}g\`{e}rement positif,
c'est-\`{a}-dire que $-\alpha $ soit remplac\'{e} par $+\alpha $ et la
spirale \textit{Amq} de rayon d\'{e}croissant est remplac\'{e} par une autre spirale
logarithmique \textit{Am'n'p'}.~\fg{} [19, p. 154]

\end{quote}

Sur la Fig. 16, ces courbes sont l\'{e}g\`{e}rement prolong\'{e}es
au-del\`{a} de $t=\frac{2\pi }{\beta }$ (un tour de cercle). L'\'{e}cart
entre une de ces deux courbes et chacun de ces cercles, est \`{a} chaque
instant $t$, \'{e}gal \`{a} la distance verticale des \'{e}l\'{e}ments
trac\'{e}s. La constante $A$ d'une courbe est d\'{e}termin\'{e}e par la
valeur de la courbe au temps $t=0$. On peut alors faire l'analyse de la m\'{e}thode de Blondel pour la
construction des oscillations entretenues de la Fig. 9 remplac\'{e}e ici par
la Fig. 17.

\begin{figure}[htbp]
\centerline{\includegraphics[width=4.73in,height=3.16in]{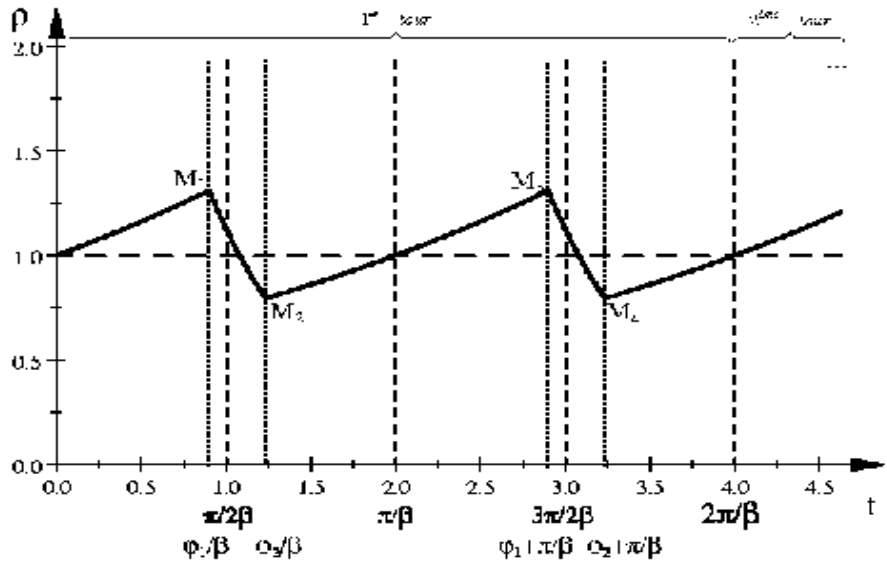}}
\caption[Repr\'{e}sentation de la Fig. 9]{Repr\'{e}sentation de la Fig. 9 dans le syst\`{e}me de coordonn\'{e}es $(t,\rho )$.}
\label{fig17}
\end{figure}

Soit $\frac{\phi _1 }{\beta }$ (resp. $\frac{\phi _2 }{\beta })$ la valeur
du temps qui correspond au point M$_{1}$ (resp. M$_{2})$. En raison des
sym\'{e}tries centrales utilis\'{e}es par Blondel pour d\'{e}finir les
points M$_{3}$ et M$_{4}$, l'abscisse de ces points vaut $\frac{\phi _1 +\pi
}{\beta }$ (resp. $\frac{\phi _2 +\pi }{\beta })$.

\newpage

\`{A} la Fig. 9, il essaye de construire une solution p\'{e}riodique (un cycle dont ne sait pas encore
s'il est attractif ou r\'{e}pulsif), de façon g\'{e}om\'{e}trique, en
s'appuyant sur le cercle C qui lui sert de guide. En partant du point A il
veut retourner au point A. Pour aller de A \`{a} M$_{1}$ (respectivement du
sym\'{e}trique de A \`{a} M$_{3})$ l'accroissement est exponentiel, le point
M$_{1}$ a pour coordonn\'{e}es $\left( {\frac{\phi _1 }{\beta }, Ae^{\left({\alpha \frac{\phi _1 }{\beta }} \right)}} \right)$. Blondel dessine ensuite
une partie de la solution de M$_{1}$ \`{a} M$_{2}$ (et de façon
sym\'{e}trique M$_{3}$ \`{a} M$_{4})$ sans donner de formule, mais son
explication est ambig\"{u}e car il dit \`{a} la fois que le r\'{e}gime
devient amorti $(\alpha <0)$ ce qui pourrait se comprendre avec les formules
d\'{e}j\`{a} utilis\'{e}e $r(t)=Ae^{+\alpha t}$ avec $\alpha $ changeant de
signe, et de façon contradictoire que \og{}le vecteur $\rho $ est diminu\'{e} d'apr\`{e}s une loi compliqu\'{e}e\fg{}. La seule contrainte
g\'{e}om\'{e}trique qu'il se donne est que la solution partant du point
M$_{2}$, doit, en croissant de la m\^{e}me façon exponentielle que de A
\`{a} M$_{1}$, passer par le sym\'{e}trique de A, puis arriver au point
M$_{3}$. Il ne calcule pas les coordonn\'{e}es du point M$_{2}$.\\

La raison probable pour laquelle Blondel est en difficult\'{e} dans cette
partie comprise entre M$_{1}$ et M$_{2}$ de la courbe, vient de l'existence
de points anguleux aux raccords en M$_{1}$, M$_{2}$, (et aussi M$_{3}$,
M$_{4})$ si l'on n'utilise que ses spirales logarithmiques comme il le
propose (ce qui est trac\'{e} \`{a} la Fig. 17). Il indique \`{a} propos
d'une figure similaire concernant les diapasons, o\`{u} la difficult\'{e}
est la m\^{e}me~:

\begin{quote}

\og{}{\ldots} le r\'{e}gime dans l'angle M$_{1}$OM$_{2}$ est troubl\'{e} et
n'est plus repr\'{e}sent\'{e} en fonction du temps que de façon
sch\'{e}matique. On conservera celle-ci en admettant que dans les angles de
r\'{e}gime troubl\'{e} le \textit{temps} varie suivant une loi diff\'{e}rente et non
explicite.~\fg{} [19, p. 155-156]

\end{quote}

\textbf{Remarques~:}\\

- Si l'on utilise strictement des spirales logarithmiques, il existe une
famille de solutions qui lient l'angle $\frac{\phi _2 }{\beta }$ \`{a}
l'ordonn\'{e}e de M$_{2}$.\\

- La th\'{e}orie des fonctions splines cr\'{e}\'{e}e en 1946, permet
actuellement de r\'{e}soudre le probl\`{e}me des points anguleux.

\newpage

Blondel, arriv\'{e} au maximum des possibilit\'{e}s d'abstraction en raison
des connaissances math\'{e}matiques de son \'{e}poque, ne s'int\'{e}resse
pas \`{a} ce qui peut se passer autour de cette solution qui pourrait
\^{e}tre un cycle limite. La diff\'{e}rence entre les Fig. 8 et 9 est
qu'effectivement le cercle C'' de la Fig. 8 pourrait \^{e}tre un cycle
attractif (on ne peut pas en sortir, sans qu'on dise comment) et C un cycle
r\'{e}pulsif, mais qu'en faisant les modifications de M$_{1}$ \`{a} M$_{2}$
et de M$_{3}$ \`{a} M$_{4}$, ce cercle C est franchi par la trajectoire et
perd son statut de cercle invariant.

Dans le syst\`{e}me de coordonn\'{e}es que nous utilisons, il est facile de
repr\'{e}senter un cycle limite attractant, en traçant autour de
l'oscillation entretenue construite \`{a} partir des spirales logarithmiques
de Blondel, des solutions partant d'un point initial proche de A, qui
convergent vers elle, comme \`{a} la Fig. 18.

\begin{figure}[htbp]
\centerline{\includegraphics[width=4.73in,height=3.16in]{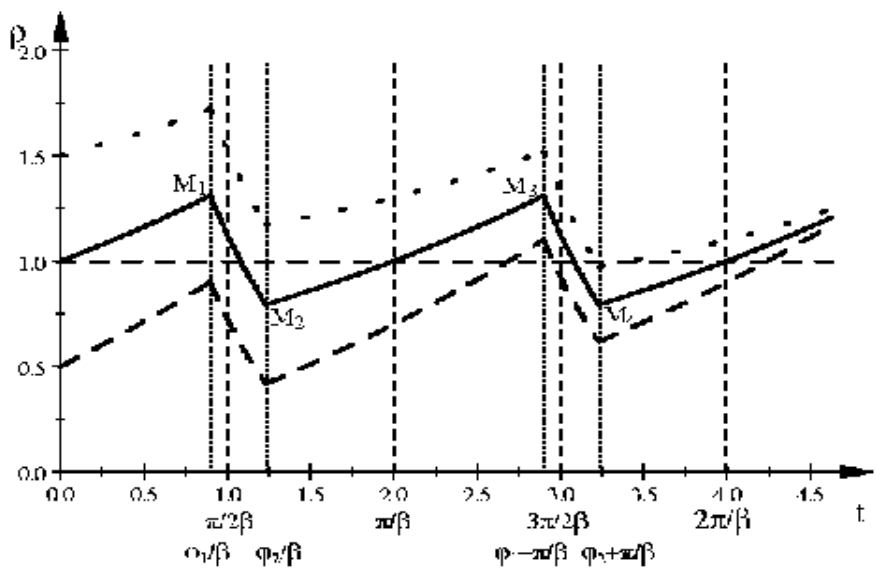}}
\caption[Solutions convergent vers le cycle limite]{Solutions convergent vers le cycle limite de la Fig. 17.}
\label{fig18}
\end{figure}

\end{document}